\documentstyle[12pt,axodraw]{article}
\catcode`@=11
\newcount\@tempcntc
\def\@citex[#1]#2{\if@filesw\immediate\write\@auxout{\string\citation{#2}}\fi
  \@tempcnta\z@\@tempcntb\m@ne\def\@citea{}\@cite{\@for\@citeb:=#2\do
    {\@ifundefined
       {b@\@citeb}{\@citeo\@tempcntb\m@ne\@citea\def\@citea{,}{\bf ?}\@warning
       {Citation `\@citeb' on page \thepage \space undefined}}%
    {\setbox\z@\hbox{\global\@tempcntc0\csname b@\@citeb\endcsname\relax}%
     \ifnum\@tempcntc=\z@ \@citeo\@tempcntb\m@ne
       \@citea\def\@citea{,}\hbox{\csname b@\@citeb\endcsname}%
     \else
      \advance\@tempcntb\@ne
      \ifnum\@tempcntb=\@tempcntc
      \else\advance\@tempcntb\m@ne\@citeo
      \@tempcnta\@tempcntc\@tempcntb\@tempcntc\fi\fi}}\@citeo}{#1}}
\def\@citeo{\ifnum\@tempcnta>\@tempcntb\else\@citea\def\@citea{,}%
  \ifnum\@tempcnta=\@tempcntb\the\@tempcnta\else
   {\advance\@tempcnta\@ne\ifnum\@tempcnta=\@tempcntb \else \def\@citea{--}\fi
    \advance\@tempcnta\m@ne\the\@tempcnta\@citea\the\@tempcntb}\fi\fi}
\catcode`@=12

\begin{document}

\begin{flushright}
RAL-TR/96--055\\
August 1996
\end{flushright}

\begin{center}
{\LARGE{\bf Generalized Pinch Technique and    }}\\[0.4cm]
{\LARGE{\bf  the Background Field Method   }}\\[0.4cm]
{\LARGE{\bf   in General Gauges  }}\\[2.5cm]
{\large Apostolos Pilaftsis}\footnote[1]{E-mail address: 
pilaftsis@mppmu.mpg.de}\footnotetext{Present address:
Max-Planck-Institute, F\"ohringer Ring 6, D-80805 M\"unchen, FRG.}\\[0.4cm]
{\em Rutherford Appleton Laboratory, Chilton, Didcot, Oxon, OX11 0QX, UK}
\end{center}
\vskip1.5cm \centerline{\bf ABSTRACT} 
It is shown that Cornwall's pinch technique can be extended in a consistent    
diagrammatic way, so as to describe general background field gauges in
Yang-Mills theories.  The resulting one-loop Green's functions are found to
obey Ward identities identical to those derived from the classical action
at the tree level.  This generalization of the pinch technique may hence be
related to the background field method implemented with novel gauge-fixing
conditions invariant under background field gauge transformations.  To one
loop, the connection between the generalized pinch technique and the
background field method in covariant and in non-covariant gauges is
explicitly demonstrated.

\newpage

\setcounter{equation}{0}
\section{Introduction}

The pinch technique (PT), as has originally been introduced by Cornwall
\cite{JMC}, is a powerful field-theoretical algorithm, which re-arranges
the $S$-matrix elements of gauge theories, such that the resulting proper
two-point, three-point, \dots, $n$-point correlation functions satisfy Ward
identities (WIs) identical to those derived from the classical Lagrangian
at the tree level \cite{JMC,BHJr,PT}.  Within this framework, the one-loop
effective PT Green's functions can further be shown to be independent
\cite{JMC,BHJr,PT,JP&AP1,KS} of the gauge-fixing conditions imposed in a
rigorous way \cite{JP&AP2}. In Ref.\ \cite{JP&AP1}, an approach has been
suggested for the construction of high-order self-energies, which are gauge
independent within the PT.  The authors of Ref.\ \cite{KP&AS} have
independently tested the gauge invariance and consistency of this approach
in a two-loop example. Apart from gauge independence, most importantly,
basic field theoretical requirements based on unitarity, analyticity and
renormalizability are satisfied for the off-shell PT correlation functions
\cite{JP&AP2}.  These conditions are deduced from resummation
considerations \cite{JP&AP1,JP&AP2}, which naturally emanate from
describing the underlying dynamics of unstable particles in spontaneous
symmetry breaking (SSB) theories, such as the Standard Model (SM) and/or
its renormalizable extensions. From that point of view, we think that a
physical meaning may be assigned to an off-shell PT Green's function 
\cite{JP&AP2}.

One may now raise the question whether the diagrammatic approach of the PT
can be formulated on the basis of the path integral, through which the
quantized Lagrangian can give rise to Green's functions exhibiting the very
same properties mentioned above.  If one relaxes the requirements of
unitarity and gauge independence, such a quantized action, for which the
derived Green's functions obey tree-level WIs, can be found with the help
of the background field method (BFM) \cite{BFM}. The focal idea of the BFM
may be explained as follows.  First, one decomposes linearly the gauge
field appearing in the classical action in terms of a background field,
$\hat{A}_\mu$, and the quantum field, $A_\mu$, which is a variable of
integration in the path integral. In the Fadeev-Popov quantization method
\cite{FP}, it is then necessary to eliminate the unphysical degrees of the
gauge field by breaking the gauge invariance of the classical Lagrangian
through a gauge-fixing condition, which is usually taken to be of covariant
form, even though such a choice of gauge fixing may not be unique. Most
importantly, the gauge-fixing condition is chosen to be invariant under
gauge transformations of the background field $\hat{A}_\mu$. Thus, the
whole Lagrangian possesses a background-field gauge invariance with respect
to the field $\hat{A}_\mu$, which only appears outside the loops. However,
the gauge symmetry is explicitly broken by the quantum field $A_\mu$, which
occurs in the loop only. In SSB theories, the latter leads to 
$\xi_Q$-dependent unphysical thresholds in the resummed off-shell
self-energies, thus spoiling the physical requirement of unitarity
\cite{JP&AP1,JP&AP2}. Only for the specific choice of the gauge-fixing
parameter $\xi_Q=1$, unitarity cuts of the one-loop Green's functions are
found to correspond to physical Landau singularities \cite{JP&AP2}. As a
consequence, the one-loop analytic results obtained by the PT coincide with
those calculated in the BFM quantized in a covariant gauge with $\xi_Q=1$
\cite{HKSY}.
       
In this paper, we shall present a different point of view. Given the above
connection between PT and BFM for $\xi_Q=1$, one may now ask the question
whether it is possible to generalize the algorithm of the PT so as to obtain
an explicit relation between the new diagrammatic approach and the BFM for
{\em any} value of $\xi_Q$.  In Section 2, we shall present an extended
version of the PT, which is here called the generalized PT (GPT), and address
the above question in the affirmative.  It is also worth emphasizing that our
generalization of the PT will be based on a gauge-dependent procedure and
so will give rise to Green's functions that will in turn depend explicitly
on the gauge fixing of the procedure chosen. Obviously, this should be
considered as a fundamental departure from the primary aim of Cornwall's
PT, which is to produce gauge-invariant Green's functions. 

After gaining some insight of the GPT in the covariant gauges in Section
2.1, we will extend our considerations into non-covariant gauges in Section
2.2, such as axial \cite{Axial,JCT1,JCT2,GL&SM,KM} or Coulomb gauges.
Again, the effective Green's functions derived with the GPT will satisfy
the usual PT or BFM WIs.  In Section 3, the Lagrangian is quantized via the
BFM in covariant and non-covariant gauges. The one-loop analytic results
obtained for the Green's functions are shown to be identical to those found
by the GPT in the corresponding gauge. The latter is demonstrated in the
scattering $q\bar{q}\to q'\bar{q}'$ in Section 4. This establishes an
explicit connection between the GPT and the BFM in a wide class of gauges.
Section 5 contains our conclusions.

\setcounter{equation}{0}
\section{Generalized pinch technique}

We shall briefly outline the main features of the PT and present the crucial
modifications pertaining to the GPT in an arbitrary gauge. We restrict
ourselves to pure Yang-Mills theories. For a comprehensive discussion on the
PT, the reader is referred to \cite{JMC_JP}.

Consider the scattering $q(p_1)\bar{q}(p_2)\to q'(k_1)\bar{q}'(k_2)$ in
the covariant $R_\xi$ gauges. The one-loop transition amplitude can 
conveniently be written down as
\begin{eqnarray}
\label{Ampl}
\langle q' \bar{q}' | T |q \bar{q} \rangle &=& 
\Gamma_\rho\Delta^{(\xi )\rho\mu} (q)\, \Pi_{\mu\nu}^{(\xi )}(q)\, 
                     \Delta^{(\xi )\nu\lambda}(q)\, \Gamma^*_\lambda\ +\ 
\Gamma_{1\mu}^{(\xi )}(q,p_1,p_2)\, \Delta^{(\xi )\mu\lambda}(q)\, 
                                  \Gamma^*_\lambda \nonumber\\
&&+\,  
\Gamma_\mu \Delta^{(\xi )\mu\lambda}(q)\, \Gamma_{1\lambda}^{(\xi )*}
(-q,k_1,k_2)\ +\ B^{(\xi )}(p_1,p_2,-k_1,-k_2)\, .
\end{eqnarray}  
Here, $p_1+p_2=-q=k_1+k_2$ and $\Gamma_\mu=(g/2)\,\bar{u}\gamma_\mu \lambda^a
v$ is the tree gluon-quark-quark vertex ($Aq\bar{q}$), where $g$ is the
strong coupling constant and $\lambda^a$ are the SU$(N)$ generators in the
fundamental representation. Furthermore, the gluon propagator in the covariant
gauges is given by 
\begin{equation}
\label{Delxi}
\Delta_{\mu\nu}^{(\xi )}(q) \ =\ \Big[\, t_{\mu\nu}(q)\, -\,
\xi \ell_{\mu\nu}(q)\, \Big]\, \frac{1}{q^2}\, ,
\end{equation}
with
\begin{equation}
\label{proj}
t_{\mu\nu}(q) \ =\ -\, g_{\mu\nu}\, +\, \frac{q_\mu q_\nu}{q^2}\, ,\qquad
\ell_{\mu\nu}(q) \ =\ \frac{q_\mu q_\nu}{q^2}\ .
\end{equation}
In Eq.\ (\ref{Ampl}), $\Pi_{\mu\nu}^{(\xi )}(q)$, $\Gamma_{1\mu}^{(\xi
  )}(q,p_1,p_2)$, $B^{(\xi )}(p_1,p_2,-k_1,-k_2)$ denote the one-loop gluon
vacuum polarization, the one-loop vertex $Aq\bar{q}$, and the box graphs,
respectively. Evidently, the total $S$-matrix element in Eq.\ (\ref{Ampl})
is gauge independent. However, the individual $\xi$ dependence present in
the one-loop correlation functions can also be eliminated with the help of
the PT diagrammatic approach \cite{PT}. Within the PT, the transition
amplitude may be recast as follows:
\begin{eqnarray}
\label{PTAmpl}
\langle q' \bar{q} '| T | q \bar{q} \rangle & = &
  \Gamma^\rho\Delta^{(\xi )}_{\rho\mu}(q)\, \widehat{\Pi}^{\mu\nu}(q)\,
             \Delta^{(\xi )}_{\nu\lambda}(q)\, \Gamma^{\lambda *}\ +\
\widehat{\Gamma}^\mu(q,p_1,p_2)\, \Delta^{(\xi )}_{\mu\lambda}(q)\, 
                                                    \Gamma^{\lambda *}
                                                                    \nonumber\\
&&+\,
\Gamma^\mu \Delta^{(\xi )}_{\mu\lambda}(q)\,
\widehat{\Gamma}^{\lambda *} (-q,k_1,k_2)\ +\ 
                                      \widehat{B}(p_1,p_2,-k_1,-k_2)\, .
\end{eqnarray} 
Correspondingly, $\widehat{\Pi}^{\mu\nu}(q)$, $\widehat{\Gamma}^\mu
(q,p_1,p_2)$ and $\widehat{B}(p_1,p_2,-k_1,-k_2)$ are the one-loop
gauge-independent PT two-, three- and four-point Green's functions. In this
approach, the effective gluon vacuum polarization turns out to be transverse,
{\em i.e.}, 
\begin{equation}
\label{Pitr}
\widehat{\Pi}_{\mu\nu}(q)\ =\ t_{\mu\nu}(q)\, \widehat{\Pi}_T (q^2)
\end{equation}
and 
\begin{equation}
\label{Wid}
q^\mu\widehat{\Gamma}_\mu (q,p_1,p_2)\ =\ 
g \Big[\, \widehat{\Sigma}(\not\! p_1)\,
                 -\, \widehat{\Sigma}(\not\! p_2 )\, \Big]\, .
\end{equation}
In Eq.\ (\ref{Wid}), $\widehat{\Sigma}(\not\!\! p)$ is the effective PT
quark self-energy, which satisfies the very same WI known from QED.
In particular, $\widehat{\Sigma}(\not\! p)$ is the usual quark self-energy
calculated in the gauge $\xi=1$, {\em i.e.}, the Feynman-'t Hooft gauge.
Because of Eqs.\ (\ref{Pitr}) and (\ref{Wid}), the remaining 
$\xi$ dependence of the tree-gluon propagators in Eq.\ (\ref{PTAmpl}) is 
trivial. Apparently, one choice dictated by simplicity would be to set 
$\xi=1$.
\vskip1.cm
\begin{center}
\begin{picture}(350,100)(0,0)
\SetWidth{0.8}
\Gluon(10,50)(50,50){3}{4}\Text(35,57)[rb]{$a,\mu, (q)$}
\LongArrow(25,43)(40,43)
\Gluon(50,50)(80,90){3}{5}\Text(80,93)[rb]{$c,\lambda, (k)$}
\LongArrow(72,70)(64.5,60)
\Gluon(80,10)(50,50){3}{5}\Text(80,10)[rt]{$b,\nu, (p)$}
\LongArrow(72,30)(64.5,40)
\Text(120,50)[]{$=$}
\BBoxc(190,50)(5,5)
\Gluon(150,50)(187.5,50){3}{4}\Text(175,57)[rb]{$a,\mu, (q)$}
\Photon(192.5,52.7)(220,90){3}{5}\Text(220,93)[rb]{$c,\lambda, (k)$}
\Photon(220,10)(192.5,47.3){3}{5}\Text(220,10)[rt]{$b,\nu, (p)$}
\Text(190,-20)[]{{\bf (a)}}
\Text(250,50)[]{$+$}
\Vertex(320,50){3}
\Gluon(280,50)(320,50){3}{4}\Text(305,57)[rb]{$a,\mu, (q)$}
\Photon(320,50)(350,90){3}{5}\Text(350,93)[rb]{$c,\lambda, (k)$}
\Photon(350,10)(320,50){3}{5}\Text(350,10)[rt]{$b,\nu, (p)$}
\Text(320,-20)[]{{\bf (b)}}
\end{picture}\\[1.cm]
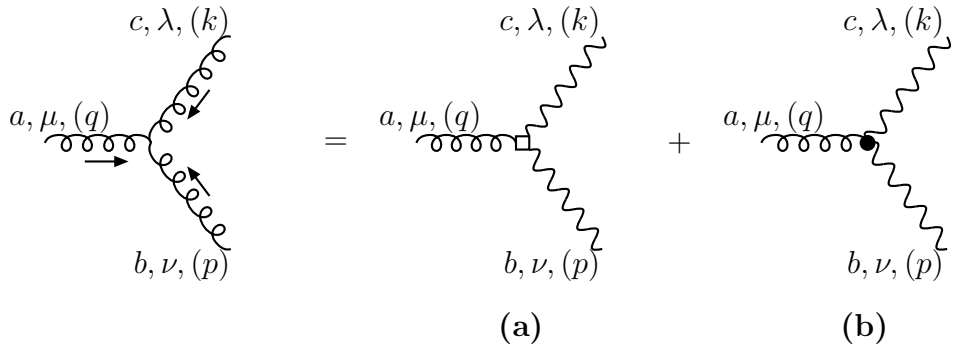
{\small {\bf Fig.\ 1:} (G)PT decomposition of the tri-gluon vertex in a general
gauge.}
\end{center}

\newpage
The PT accomplishes to arrive at Eq.\ (\ref{PTAmpl}) starting from 
Eq.\ (\ref{Ampl}). The observation to be made is that the residual 
$\xi$ dependence of $\Pi^{(\xi )}_{\mu\nu}(q)$, for example, is hidden in the
vertices and boxes and should therefore be ``pinched'' out. These ``pinching''
(or PT) terms are kinematically indistinguishable with that of 
$\Pi^{(\xi )}_{\mu\nu}(q)$ and should therefore be added to it, yielding the
gauge-independent PT self-energy $\widehat{\Pi}_{\mu\nu}(q)$. Similarly, the
PT effective vertex $\widehat{\Gamma}_\mu (q,p_1,p_2)$ receives PT
contributions from the box graphs. 

In order to extract the PT terms, one has to employ some elementary
``WIs'' at the $S$-matrix level. Each time a loop momentum $k^\mu$ gets
contracted with a $\gamma_\mu$ of an internal quark line, it gives
rise to identities of the kind
\begin{equation}
\label{EWI}
\not\! k\ =\ (\not\! k_1\, +\, \not\! k\, -\, m_q)\ -\ ( \not\! k_1\, -\, 
m_q)\, . 
\end{equation}
In general, such pinching momenta can originate either from the
$k^\mu$-dependent part of the gluon propagator in the loop or from the
tri-gluon vertex depicted in Fig.\ 1. This is diagrammatically shown in
Fig.\ 2.  In a typical gauge-dependent vertex graph, the first term of the
RHS of Eq.\ (\ref{EWI}) cancels the virtual fermion propagator, whereas the
second one vanishes for external on-shell quarks.  This algorithm produces
self-energy-type graphs, which we call self-energy PT parts, shown in Fig.\ 
2(b). These self-energy PT terms should be allotted to the proper two-point
correlation function evaluated in the given gauge. In this context, we must
remark that the gauge dependence in the QED-like part of $\Gamma_\mu^{(\xi
  )}(q)$ vanishes identically. As can be seen from Fig.\ 3 in a
diagrammatic manner, even for a propagator in the axial gauge \cite{Axial}
\begin{equation}
\label{Delaxial}
\Delta^{(\eta )}_{\mu\nu} (q)\ =\ \Big[\, -\, g_{\mu\nu}\, +\, 
\frac{q_\mu \eta_\nu + q_\nu \eta_\mu}{q\cdot \eta}\, -\,
\frac{\eta^2 q_\mu q_\nu}{(q\cdot\eta)^2}\, \Big]\, \frac{1}{q^2}\, ,
\end{equation}
the terms not proportional to $g_{\mu\nu}$ and $\eta^2$ produce PT terms that
add to zero in the dimensional regularization (DR). Without loss of
generality, we will adopt DR in the following, since massless tadpole
integrals do not contribute in this scheme. 

\begin{center}
\begin{picture}(400,100)(0,0)
\SetWidth{0.8}
\ArrowLine(0,80)(30,50)\Text(0,89)[l]{$q(p_1)$}
\ArrowLine(30,50)(0,20)\Text(0,11)[l]{$\bar{q}(p_2)$}
\Gluon(30,50)(60,50){3}{3}
\Gluon(60,50)(90,80){3}{4}
\Gluon(90,20)(60,50){3}{4}
\ArrowLine(90,20)(90,80)
\ArrowLine(90,80)(120,80)\Text(127,89)[r]{$q'(k_1)$}
\ArrowLine(120,20)(90,20)\Text(127,11)[r]{$\bar{q}'(k_2)$}

\Text(130,50)[]{$=$}

\ArrowLine(140,80)(170,50)
\ArrowLine(170,50)(140,20)
\BBoxc(200,50)(5,5)
\Gluon(170,50)(197.5,50){3}{3}
\Photon(202.5,52.7)(230,80){3}{3.6}
\Photon(230,20)(202.5,47.3){3}{3.6}
\ArrowLine(230,20)(230,80)
\ArrowLine(230,80)(260,80)
\ArrowLine(260,20)(230,20)
\Text(200,-10)[]{{\bf (a)}}

\Text(270,50)[]{$+$}

\ArrowLine(280,80)(310,50)
\ArrowLine(310,50)(280,20)
\Gluon(310,50)(340,50){3}{3}
\PhotonArc(360,50)(20,0,180){3}{6.2}
\PhotonArc(360,50)(20,180,360){3}{6}
\Vertex(340,50){3}
\ArrowLine(380,50)(410,80)
\ArrowLine(410,20)(380,50)\Vertex(380,50){1.7}
\Text(340,-10)[]{{\bf (b)}}
\end{picture}\\[0.7cm]
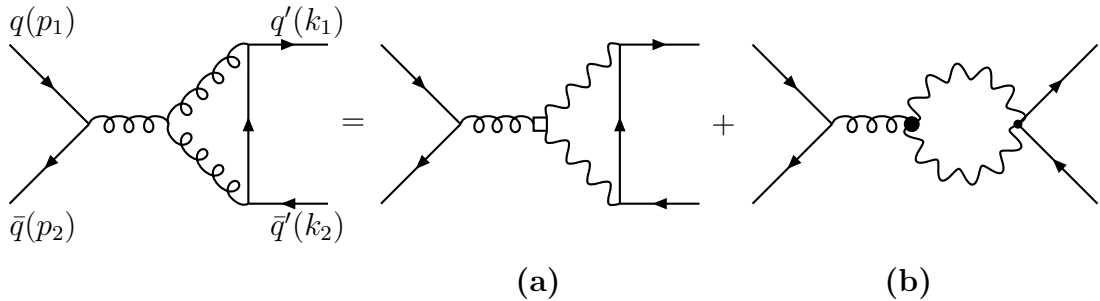
{\small {\bf Fig.\ 2:} One-loop (G)PT vertex (a) and its propagator-like 
counterpart (b).}
\end{center}

\begin{center}
\begin{picture}(400,300)(0,0)
\SetWidth{0.8}
\ArrowLine(0,290)(40,250)
\ArrowLine(40,250)(0,210)
\Gluon(40,250)(70,250){3}{3}
\ArrowLine(70,250)(95,275)\ArrowLine(95,275)(110,290)
\ArrowLine(110,210)(95,225)\ArrowLine(95,225)(70,250)
\Gluon(95,225)(95,275){-3}{4}

\Text(120,250)[]{$+$}

\ArrowLine(130,290)(170,250)
\ArrowLine(170,250)(130,210)
\Gluon(170,250)(200,250){3}{3}
\ArrowLine(200,250)(210,260)\ArrowLine(210,260)(230,280)
                                                  \ArrowLine(230,280)(240,290)
\ArrowLine(240,210)(200,250)
\GlueArc(220,270)(10,-135,45){2.7}{5}

\Text(250,250)[]{$+$}

\ArrowLine(260,290)(300,250)
\ArrowLine(300,250)(260,210)
\Gluon(300,250)(330,250){3}{3}
\ArrowLine(330,250)(370,290)
\ArrowLine(370,210)(360,220)\ArrowLine(360,220)(340,240)
                                                  \ArrowLine(340,240)(330,250)
\GlueArc(350,230)(10,-45,135){2.7}{5}

\LongArrow(0,150)(20,150)\Text(10,160)[]{{\small (G)PT}}

\ArrowLine(40,180)(70,150)
\ArrowLine(70,150)(40,120)
\Gluon(70,150)(100,150){3}{3}\Vertex(100,150){1.7}
\ArrowLine(100,150)(130,180)
\ArrowLine(130,120)(120,130)\ArrowLine(120,130)(100,150)
\PhotonArc(110,140)(11,-45,135){2.7}{5.5}

\Text(200,150)[]{$-$}

\ArrowLine(240,180)(270,150)
\ArrowLine(270,150)(240,120)
\Gluon(270,150)(300,150){3}{3}\Vertex(300,150){1.7}
\ArrowLine(300,150)(320,170)\ArrowLine(320,170)(330,180)
\ArrowLine(330,120)(300,150)
\PhotonArc(310,160)(11,-135,45){2.7}{5.5}

\Text(10,50)[]{$+$}

\ArrowLine(40,80)(70,50)
\ArrowLine(70,50)(40,20)
\Gluon(70,50)(100,50){3}{3}\Vertex(100,50){1.7}
\ArrowLine(100,50)(120,70)\ArrowLine(120,70)(130,80)
\ArrowLine(130,20)(100,50)
\PhotonArc(110,60)(11,-135,45){2.7}{5.5}

\Text(200,50)[]{$-$}

\ArrowLine(240,80)(270,50)
\ArrowLine(270,50)(240,20)
\Gluon(270,50)(300,50){3}{3}\Vertex(300,50){1.7}
\ArrowLine(300,50)(330,80)
\ArrowLine(330,20)(320,30)\ArrowLine(320,30)(300,50)
\PhotonArc(310,40)(11,-45,135){2.7}{5.5}

\end{picture}\\
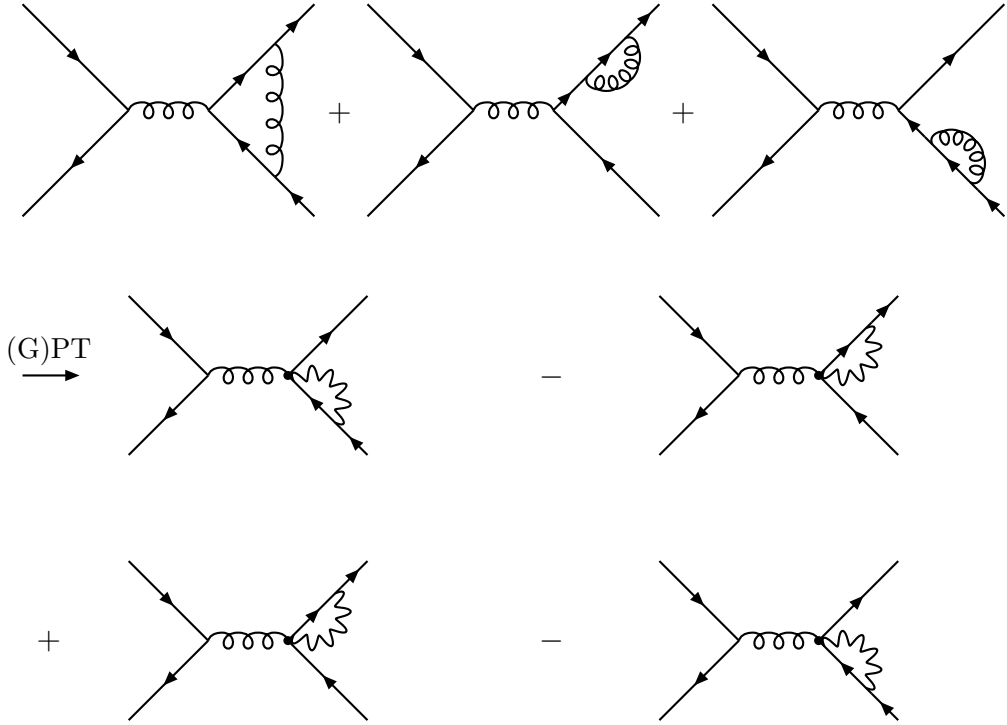
{\small {\bf Fig.\ 3:} Vanishing of the (G)PT terms in the axial gauge for
the Abelian part of $\Gamma_{1\mu} (q,p_1,p_2)$.}
\end{center}

Apart from self-energy PT terms originating from vertex graphs, a box diagram 
can also contribute PT terms to $\widehat{\Pi}_{\mu\nu} (q)$ for $\xi\not= 1$,
as has been displayed in Fig.\ 4. Since the PT self-energy is independent of
the gauge-fixing parameter $\xi$ \cite{JMC_JP}, the Feynman-'t Hooft gauge
provides a great computational simplification. In this gauge, the only pinching
momenta are contained in the tri-gluon vertex, 
\begin{equation}
\label{3g}
\Gamma_{\mu\nu\lambda}^{abc}(q,p,k)\ =\ gf^{abc}\Big[\,
(p-k)_\mu g_{\nu\lambda}  \, +\, (k-q)_\nu g_{\mu\lambda} \, +\,
(q-p)_\lambda g_{\mu\nu} \, \Big]\, ,
\end{equation}
where $q+p+k=0$ and all four-momenta are incoming, as shown in Fig.\ 1. 
To make this explicit, we first decompose
$f^{abc}\Gamma_{\mu\nu\lambda}(q,p,k)$ in the following way: 
\begin{equation}
\label{3gdecomp}
\Gamma_{\mu\nu\lambda}(q,p,k)\ =\ \Gamma^F_{\mu\nu\lambda}(q,p,k)\,
+\, \Gamma^P_{\mu\nu\lambda}(q,p,k)\, ,
\end{equation}
with
\begin{eqnarray}
\label{3gF}
\Gamma^F_{\mu\nu\lambda}(q,p,k) & = & g \Big[\, (p-k)_\mu g_{\nu\lambda}\, 
-\, 2q_\nu g_{\mu\lambda}\, +\, 2q_\lambda g_{\mu\nu}\, \Big]\, ,\\
\label{3gP}
\Gamma^P_{\mu\nu\lambda}(q,p,k) & = & g\, (  k_\lambda g_{\mu\nu}\,   
-\,  p_\nu g_{\mu\lambda} )\, .
\end{eqnarray}
Notice that the splitting of the tri-gauge coupling in Eq.\ (\ref{3gdecomp})
makes reference to the external gluon, $A^a_\mu$ say. This explicitly breaks
the cyclic symmetry of $\Gamma_{\mu\nu\lambda}(q,p,k)$, which is not
present neither in $\Gamma^F_{\mu\nu\lambda}(q,p,k)$ nor in
$\Gamma^P_{\mu\nu\lambda}(q,p,k)$. To underline this feature, we shall
diagrammatically represent the gluons in the loop by wavy lines, {\em e.g.},
see Figs.\ 2(a) and 2(b).  It is now easy to recognize that
$\Gamma^P_{\mu\nu\lambda}$ in Eq.\ (\ref{3gP}) contains the pinching loop
momenta $k$ and $p$. In addition, we have
\begin{eqnarray}
\label{3gWI}
\frac{1}{g}\, q^\mu\, \Gamma_{\mu\nu\lambda} (q,p,k) & =& 
                t_{\nu\lambda}(p)\, p^2\  -\ t_{\nu\lambda}(k)\, k^2\nonumber\\
        &=& U^{-1}_{\nu\lambda}(p)\, -\, U^{-1}_{\nu\lambda}(k)\, ,\\
\label{3gWIF}
\frac{1}{g}\, q^\mu\, \Gamma^F_{\mu\nu\lambda} (q,p,k) & =&
                       -g_{\nu\lambda} p^2\, +\, g_{\nu\lambda}  k^2\nonumber\\
         &=& \Delta^{F-1}_{\nu\lambda}(p)\, -\, 
                                              \Delta^{F-1}_{\nu\lambda}(k)\, ,
\end{eqnarray}
where $U^{-1}_{\mu\nu}(q)=t_{\mu\nu}(q)\, q^2$ and
$\Delta^{F-1}_{\mu\nu}(q)=-g_{\mu\nu}q^2$ are the inverse gluon propagators
in the unitary and the $\xi = 1$ gauge, respectively. Note that
$U^{-1}_{\nu\lambda}(q)$ does not have any inverse, unless a fictitious SSB
mass is introduced for the massless gluon in order to cope with infra-red
(IR) infinities.

\begin{center}
\begin{picture}(370,100)(0,0)
\SetWidth{0.8}

\ArrowLine(0,80)(30,80)\Text(0,90)[l]{$q(p_1)$}
\ArrowLine(30,20)(0,20)
\ArrowLine(30,80)(30,20)\Text(0,10)[l]{$\bar{q}(p_2)$}
\Gluon(30,80)(70,80){3}{3}
\Gluon(30,20)(70,20){-3}{3}
\ArrowLine(70,80)(100,80)\Text(110,90)[r]{$q'(k_1)$}
\ArrowLine(100,20)(70,20)\Text(110,10)[r]{$\bar{q}'(k_2)$}
\ArrowLine(70,20)(70,80)

\Text(120,50)[]{$=$}

\ArrowLine(140,80)(170,80)
\ArrowLine(170,20)(140,20)
\ArrowLine(170,80)(170,20)
\Photon(170,80)(210,80){3}{4}
\Photon(170,20)(210,20){-3}{4}
\ArrowLine(210,80)(240,80)
\ArrowLine(240,20)(210,20)
\ArrowLine(210,20)(210,80)
\Text(195,-10)[]{{\bf (a)}}

\Text(250,50)[]{$+$}

\ArrowLine(270,80)(300,50)
\ArrowLine(300,50)(270,20)
\PhotonArc(320,50)(20,0,180){3}{6.2}\Vertex(300,50){1.7}
\PhotonArc(320,50)(20,180,360){3}{6}\Vertex(340,50){1.7}
\ArrowLine(340,50)(370,80)
\ArrowLine(370,20)(340,50)
\Text(320,-10)[]{{\bf (b)}}

\end{picture}\\[0.7cm]
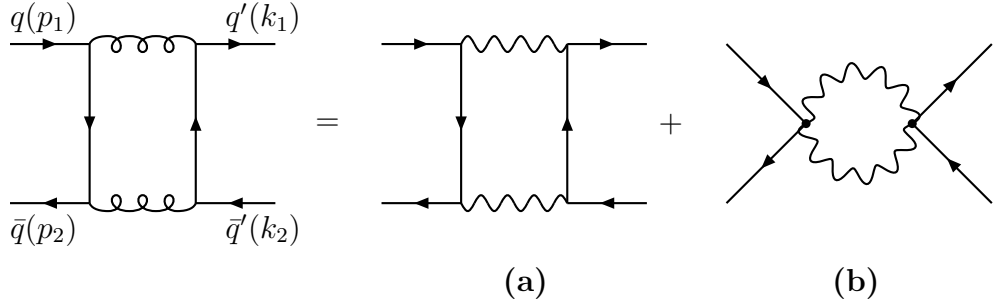
{\small {\bf Fig.\ 4:} (G)PT decomposition of the box graph.}
\end{center}

It is important to notice that the decomposition of $\Gamma_{\mu\nu\lambda} 
(q,p,k)$ in Eq.\ (\ref{3gdecomp}) will imply the transversality of
$\widehat{\Pi}_{\mu\nu}(q)$. After extracting the self-energy PT terms induced
by $\Gamma^P_{\mu\nu\lambda} (q,p,k)$ and considering the ghost loop, one
obtains the analytic expression
\cite{JMC_JP} 
\begin{equation}
\label{PTQCD}
\widehat{\Pi}_{\mu\nu} (q)\ =\ 
       \frac{1}{4}c_A\, \int\, \frac{d^nk}{i(2\pi )^n}\, \frac{1}{k^2\, p^2}\, 
\Big[\Gamma^F_{\mu\lambda\sigma}(q,p,k)\Gamma_\nu^{F\lambda\sigma}(q,p,k)\, 
-\, 2S^F_\mu(p,k) S^F_\nu (p,k)\, \Big]\, ,
\end{equation}
where $n=4-2\varepsilon$, $p=-q-k$, $c_A=N$ is the Casimir factor in the
adjoint representation of SU$(N)$, and 
\begin{equation}
\label{PTghost}
S^F_\mu (p,k)\ =\ g\, (p-k)_\mu\, . 
\end{equation}
It is now straightforward to check that indeed,
\begin{equation}
\label{QCDtrans}
q^\mu \widehat{\Pi}_{\mu\nu} (q)\ =\ 0\, ,
\end{equation} 
on account of Eq.\ (\ref{3gWIF}) and the fact that massless tadpoles vanish
in DR${}$. Furthermore, it has been noticed \cite{HKSY} that the result
given in Eq.\ (\ref{PTQCD}) is identical to that obtained by the BFM for
the gauge-fixing parameter value $\xi_Q=1$ in a covariant gauge condition
\cite{BFM}. Note that this gauge-fixing condition (see Eq.\ (\ref{BFMxiQ})
below) is different from the usual one imposed in $R_\xi$ gauges.  In
particular, it is easy to recognize \cite{Abbott} that for $\xi_Q=1$, the
coupling of the gluon background field, $\hat{A}$, to quantum gluons, $A$,
and ghosts, $c_g$, are equal to $f^{abc}\Gamma^F_{\mu\nu\lambda}(q,p,k)$
and $f^{abc} S^F_\mu(p,k)$, respectively. We will return to this point in
Section 3.

In the derivation of Eq.\ (\ref{QCDtrans}), the WI in Eq.\ (\ref{3gWIF})
has been crucial. One could therefore ask the question whether the PT can
be generalized by modifying the conventional decomposition of Eq.\ 
(\ref{3gdecomp}), so that an elementary Abelian-type WI analogous to Eq.\ 
(\ref{3gWIF}) is satisfied. Suppose we make the decomposition of the
tri-gluon vertex,
\begin{equation}
\label{xiQGPT}
\Gamma_{\mu\nu\lambda}(q,p,k)\ =\ \Gamma^{(\xi_Q )}_{\mu\nu\lambda} (q,p,k)\ 
+\ \Gamma^{P(\xi_Q )}_{\mu\nu\lambda}(q,p,k)\, ,
\end{equation} 
for a fixed given $\xi$, $\xi=\xi_Q$ say, such that
\begin{equation}
\label{xiQWI}
\frac{1}{g}\, q^\mu\, \Gamma^{(\xi_Q )}_{\mu\nu\lambda}(q,p,k)\ =\ 
\Delta^{(\xi_Q )-1}_{\nu\lambda}(p)\, -\, 
                         \Delta^{(\xi_Q )-1}_{\nu\lambda}(k)\, ,
\end{equation}
with
\begin{equation}
\label{invDelxi}
\Delta^{(\xi_Q )-1}_{\mu\nu}(q)\ =\ q^2\, \Big[ t_{\mu\nu}(q)\, -\, 
\frac{1}{\xi_Q}\, \ell_{\mu\nu}(q) \Big]\, .
\end{equation}
Moreover, we require that Cornwall's PT be recovered for $\xi_Q =
1$.\footnote[1]{The decomposition in Eq.\ (\ref{xiQGPT}) may be similar to
  that given by the author in Ref.\ \cite{BHJr}.  His motivation was to
  develop an ultra-violet-improved gauge technique by including a
  non-perturbative transverse $Aq\bar{q}$ vertex in the Schwinger--Dyson
  equation for the quark self-energy. However, our main interest and
  theoretical analysis are very different from Ref.\ \cite{BHJr}.} Then,
pinching momenta will arise from $\Gamma^{P(\xi_Q)}_{\mu\nu\lambda}(q,p,k)$
and from the difference of propagators $\Delta^{(\xi)}_{\mu\nu } (q) -
\Delta^{(\xi_Q)}_{\mu\nu } (q) = \ell_{\mu\nu}(q)\, (\xi_Q -\xi )/q^2$.
This generalized version of the PT, the GPT, will be formulated in the
covariant $R_\xi$ gauges in Section 2.1 and in non-covariant gauges in
Section 2.2.

\subsection{GPT in covariant gauges}

In this section, we shall formulate the GPT in the covariant $R_\xi$ gauges
and argue that this extended version of the PT has very similar features
with Cornwall's PT \cite{JMC}.

We start again from the decomposition in Eq.\ (\ref{xiQGPT}),
\begin{displaymath}
\Gamma_{\mu\nu\lambda}(q,p,k)\ =\ \Gamma^{(\xi_Q )}_{\mu\nu\lambda} (q,p,k)\
+\ \Gamma^{P(\xi_Q )}_{\mu\nu\lambda}(q,p,k)\, .
\end{displaymath}
It is not difficult to find that the term 
$\Gamma^{(\xi_Q )}_{\mu\nu\lambda} (q,p,k)$ in compliance with the WI of Eq.\ 
(\ref{xiQWI}) may be given by
\begin{eqnarray}
\label{3gxiQ}
\Gamma^{(\xi_Q )}_{\mu\nu\lambda} (q,p,k) 
                                    & =& g \Big[\, (p-k)_\mu g_{\nu\lambda}\, 
-\, 2q_\nu g_{\mu\lambda}\, +\, 2q_\lambda g_{\mu\nu}\, \nonumber\\
&&+\, \Big(1-\frac{1}{\xi_Q}\Big) k_\lambda g_{\mu\nu}\, -\, 
\Big(1-\frac{1}{\xi_Q}\Big)p_\nu g_{\mu\lambda}\, \Big]\, ,
\end{eqnarray}
and 
\begin{equation}
\label{3gxiQP}
\Gamma^{P(\xi_Q )}_{\mu\nu\lambda}(q,p,k)\ =\ g\, \frac{1}{\xi_Q}\, \Big( 
k_\lambda g_{\mu\nu}\, -\, p_\nu g_{\mu\lambda}\, \Big)\, .
\end{equation}
Moreover, the propagator $\Delta^{(\xi_Q )}_{\mu\nu} (q)$ is defined in Eq.\
(\ref{Delxi}) for $\xi=\xi_Q$ and its inverse in Eq.\ (\ref{invDelxi}). Note
that the traditional PT \cite{JMC} is restored in the limit $\xi_Q\to 1$, in
which Eqs.\ (\ref{3gxiQ}) and (\ref{3gxiQP}) collapse correspondingly to Eqs.\
(\ref{3gF}) and (\ref{3gP}). In the same limit, the gluon propagator
$\Delta^{(\xi_Q )}_{\mu\nu}(q )$ goes into $\Delta^F_{\mu\nu}(q)$.

We will now see how GPT operates at the $S$-matrix level, having features
very similar to those of the PT \cite{JMC}. First, we must observe that 
$\Gamma^{P(\xi_Q )}_{\mu\nu\lambda}(q,p,k)$ indeed ``pinches" in Fig.\ 2.
For simplicity, we consider that the gluon propagators are in the gauge
$\xi =\xi_Q$.  Then, the only pinching contributions in Fig.\ 2 come 
from the expression
\begin{equation}
\label{xiQGamP}
\frac{1}{g}\, \Gamma^{P(\xi_Q )}_{\mu\nu\lambda} (q,p,k)\, 
\Delta_\rho^{(\xi_Q )\nu}(p)\, \Delta^{(\xi_Q )\lambda}_\sigma (k)\ =\
-\, \frac{k_\sigma}{k^2}\, \Delta^{(\xi_Q )}_{\mu\rho} (p)\, +\,
\frac{p_\rho}{p^2}\, \Delta^{(\xi_Q )}_{\mu\sigma} (k)\, .
\end{equation}
Indeed, the RHS of Eq.\ (\ref{xiQGamP}) has the correct structure to provide
the self-energy GPT terms for a given $\xi_Q$ by means of Eq. (\ref{EWI}).
Again, setting $\xi_Q=1$ in Eq.\ (\ref{xiQGamP}), Cornwall's pinching 
procedure is fully recovered. The remaining tri-gauge coupling 
$\Gamma^{(\xi_Q )}_{\mu\nu\lambda}(q,p,k)$, together with the QED-like 
graphs shown in Fig.\ 3, give rise to the one-loop effective GPT 
$Aq\bar{q}$-coupling, denoted by $\widehat{\Gamma}^{(\xi_Q )}_\mu(q,p_1,p_2)$.
In addition, we have
\begin{equation}
\label{xiQWid}
q^\mu\, \widehat{\Gamma}^{(\xi_Q )}_\mu (q,p_1,p_2)\ =\ 
g \Big[\, \widehat{\Sigma}^{(\xi_Q )} (\not\! p_1)\,
                 -\, \widehat{\Sigma}^{(\xi_Q )} (\not\! p_2 )\, \Big]\, .
\end{equation}
which is exactly the WI of Eq.\ (\ref{Wid}). Correspondingly,
$\widehat{\Sigma}^{(\xi_Q )}(\not\! p)$ is the GPT quark self-energy, which
coincides with the usual quark self-energy evaluated in the gauge $\xi=\xi_Q$.
After all the self-energy GPT terms induced by the {\em generalized} pinching
momenta in Eq.\ (\ref{xiQGamP}) have been identified and added to 
$\Pi^{(\xi_Q )}_{\mu\nu} (q)$, as shown in Fig.\ 5, the effective GPT
self-energy, $\widehat{\Pi}^{(\xi_Q )}_{\mu\nu} (q)$, takes on the analytic
form
\begin{eqnarray}
\label{xiQPTQCD}
\widehat{\Pi}^{(\xi_Q)}_{\mu\nu} (q) & = & 
       \frac{c_A}{4}\, \int\, \frac{d^nk}{i(2\pi )^n}\,
\Big[\, \Delta^{(\xi_Q )}(p) \Delta^{(\xi_Q )}(k)\, 
\Gamma^{(\xi_Q)}_\mu(q,p,k)\Gamma_\nu^{(\xi_Q)}(q,p,k)\nonumber\\
&& - \, \frac{2}{k^2p^2}\, S^F_\mu(p,k) S^F_\nu (p,k)\, \Big]\, ,
\end{eqnarray}
where all contracted Lorentz indices on the RHS of Eq.\ (\ref{xiQPTQCD})
are not explicitly displayed. At this stage, one can already see the 
connection between the GPT and the BFM in the covariant gauges, when 
comparing $\widehat{\Pi}^{(\xi_Q )}_{\mu\nu}(q)$ with the BFM gluon
self-energy given in Ref.\ \cite{HKSY,Abbott}. This relation will be further
elaborated in Section 4. 

\begin{center}
\begin{picture}(410,300)(0,0)
\SetWidth{0.8}

\ArrowLine(0,280)(30,250)
\ArrowLine(30,250)(0,220)
\Gluon(30,250)(50,250){3}{2}
\GlueArc(65,250)(15,0,180){3}{5}
\GlueArc(65,250)(15,180,360){3}{5}
\Gluon(80,250)(100,250){3}{2}
\ArrowLine(100,250)(130,280)
\ArrowLine(130,220)(100,250)

\Text(140,250)[]{$+$}

\ArrowLine(150,280)(180,250)
\ArrowLine(180,250)(150,220)
\Gluon(180,250)(200,250){3}{2}
\DashArrowArc(215,250)(15,0,180){1}
\DashArrowArc(215,250)(15,180,360){1}
\Gluon(230,250)(250,250){3}{2}
\ArrowLine(250,250)(280,280)
\ArrowLine(280,220)(250,250)

\Text(290,250)[]{$+$}

\ArrowLine(300,280)(330,250)
\ArrowLine(330,250)(300,220)
\Gluon(330,250)(350,250){3}{2}
\PhotonArc(365,250)(15,0,180){3}{6.2}
\PhotonArc(365,250)(15,180,360){3}{6}
\Vertex(350,250){3}
\ArrowLine(380,250)(410,280)
\ArrowLine(410,220)(380,250)\Vertex(380,250){1.7}

\Text(10,150)[]{$+$}

\ArrowLine(40,180)(70,150)
\ArrowLine(70,150)(40,120)
\PhotonArc(85,150)(15,0,180){3}{6.2}
\PhotonArc(85,150)(15,180,360){3}{6}
\Gluon(100,150)(120,150){3}{2}
\Vertex(100,150){3}
\ArrowLine(120,150)(150,180)
\ArrowLine(150,120)(120,150)\Vertex(70,150){1.7}
   
\Text(175,150)[]{$+$}

\ArrowLine(200,180)(230,150)
\ArrowLine(230,150)(200,120)
\PhotonArc(250,150)(20,0,180){3}{6.2}\Vertex(230,150){1.7}
\PhotonArc(250,150)(20,180,360){3}{6}\Vertex(270,150){1.7}
\ArrowLine(270,150)(300,180)
\ArrowLine(300,120)(270,150)

\Text(10,50)[]{$=$}

\ArrowLine(40,80)(70,50)
\ArrowLine(70,50)(40,20)
\Gluon(70,50)(88.5,50){3}{2}\Text(80,55)[b]{$\hat{A}$}
\BBoxc(90,50)(5,5)
\PhotonArc(105,50)(15,6,171){3}{5}\Text(105,70)[b]{$A$}
\PhotonArc(105,50)(15,189,354){3}{5}\Text(105,30)[t]{$A$}
\BBoxc(120,50)(5,5)
\Gluon(122.5,50)(140,50){3}{2}\Text(130,55)[b]{$\hat{A}$}
\ArrowLine(140,50)(170,80)
\ArrowLine(170,20)(140,50)

\Text(200,50)[]{$+$}

\ArrowLine(230,80)(260,50)
\ArrowLine(260,50)(230,20)
\Gluon(260,50)(278.5,50){3}{2}\Text(270,55)[b]{$\hat{A}$}
\BBoxc(280,50)(5,5)
\DashArrowArc(295,50)(15,6,171){1}\Text(295,73)[b]{$c_g$}
\DashArrowArc(295,50)(15,189,354){1}\Text(295,30)[t]{$c_g$}
\BBoxc(310,50)(5,5)
\Gluon(312.5,50)(330,50){3}{2}\Text(320,55)[b]{$\hat{A}$}
\ArrowLine(330,50)(360,80)
\ArrowLine(360,20)(330,50)

\end{picture}\\[0.7cm]
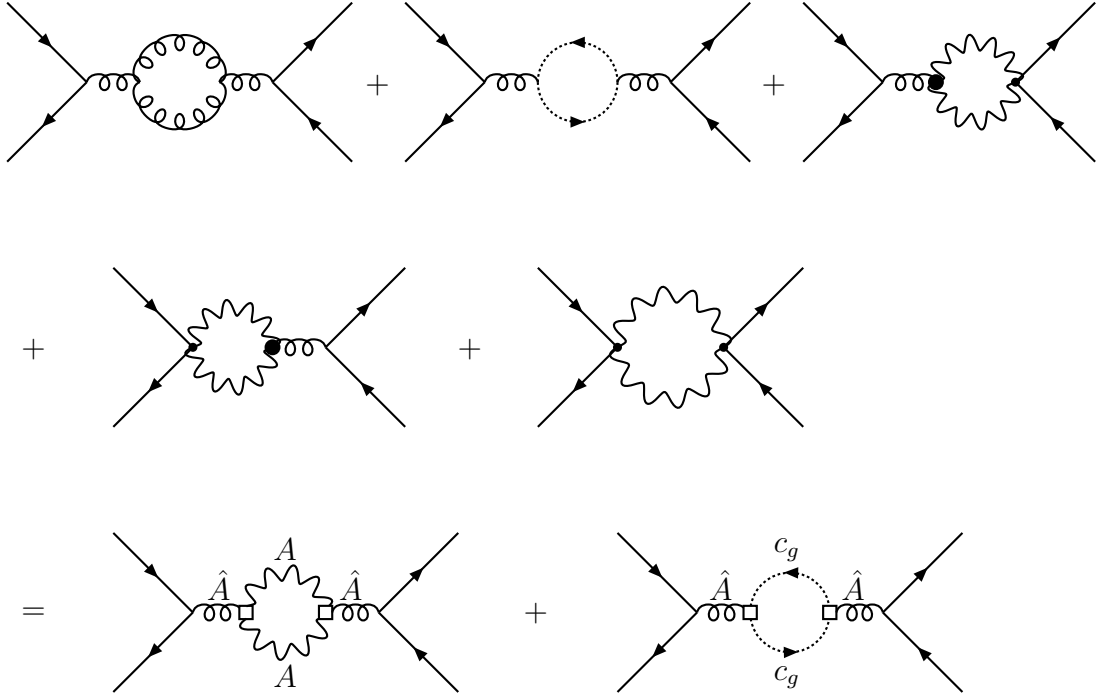
{\small {\bf Fig.\ 5:} Connection between GPT and BFM in a $\xi=\xi_Q$ gauge
for the self-energy $\widehat{\Pi}^{(\xi_Q )}_{\mu\nu}$.}
\end{center}

Even though we have worked in a gauge, in which the virtual gluon
propagators have been gauge-fixed in $\xi=\xi_Q$, one can, however, check
that our results would have remained unaffected if we had chosen another
gauge. The algorithm of the GPT in the $\xi_Q$ gauge is completely
specified, as long as the steps contained in Eqs.\ 
(\ref{xiQGPT})--(\ref{3gxiQP}) are explicitly given. In fact, these would
not change, even if the gluon propagators were taken in the axial gauge
given in Eq.\ (\ref{Delaxial}).  In this case, it is important to identify
what the {\em generalized} pinching momenta are. These GPT momenta
originate from the gluon propagators difference
\begin{equation}
\label{xiQdiff}
\Delta^{(\eta )}_{\mu\nu} (q)\, -\, \Delta^{(\xi_Q )}_{\mu\nu} (q) \ =\
\Big[\, \Big(\xi_Q - \frac{\eta^2 q^2}{(q\cdot\eta )^2} \Big)
\frac{q_\mu q_\nu}{q^2}\, +\, \frac{q_\mu\eta_\nu +
q_\nu\eta_\mu}{q\cdot\eta}\, \Big]\, \frac{1}{q^2} 
\end{equation}
and the tri-gauge generalized pinching part of the vertex,
$\Gamma^{P(\xi_Q)}_{\mu\nu\lambda}$. In particular, the process \cite{NJW} and
gauge independence \cite{KS} of the GPT may be shown rigorously by virtue of
Becchi-Rouet-Stora identities \cite{BRS}, as was done in Ref.\ \cite{JP&AP2}.
We will not pursue this topic here any further. Instead, we shall apply the
GPT in non-covariant gauges. 

\subsection{GPT in non-covariant gauges}

Following the method developed in the previous section, we shall extend the
GPT beyond the covariant gauges, {\em i.e.}, the non-covariant gauges
\cite{Axial,JCT1,JCT2}. Let us first consider the general non-covariant
gauge-fixing condition \cite{JCT1}
\begin{equation}
\label{etaGF}
G^a [A]\ =\ \frac{\eta^\mu\eta^\nu}{\eta^2}\, \partial_\mu A^a_\nu\, .
\end{equation}
The above condition enters the quantum Lagrangian via the gauge-fixing
term
\begin{equation}
\label{etaGFL}
{\cal L}_\eta\ =\ -\, \frac{1}{2 \xi (\eta^2)^2}\, (\eta^\mu\eta^\nu
\partial_\mu A^a_\nu)^2\, ,
\end{equation}
where $\eta^\mu$ is an arbitrary but constant four-vector. In general, we can 
classify the non-covariant gauges from the different values of $\eta^2$, {\em
i.e.}, $\eta^2<0$ (axial gauge), $\eta^2=0$ (light-cone gauge), $\eta^2>0$
(Hamilton or time-like gauge). For $\xi =0$, one may use Lagrange multipliers
$L^a$ and write ${\cal L}_\eta$ as 
\begin{equation}
\label{etaLM}
{\cal L}_\eta\ =\ -\, L^a\, (\eta^\mu \eta^\nu \partial_\mu A^a_\nu)\, ,
\end{equation}
where $L^a$ is an auxiliary field that mixes with the gluon $A^a$. This
leads to a proliferation of Feynman rules. Therefore, it may be more
convenient to work with $\xi\not=0$ and then take the limit $\xi \to 0$.
In this limit, the gauge-fixing condition (\ref{etaGF}) leads to the gluon
propagator in Eq.\ (\ref{Delaxial}).  To avoid excessive complication, we
set $\xi =1$ in the following, unless it is explicitly stated otherwise.
The latter, however, does not confine the generality of our formulation
concerning the GPT in the non-covariant gauges.

Considering the gauge-fixing term in Eq.\ (\ref{etaGFL}), the inverse
propagator is written down
\begin{equation}
\label{etainvDel}
\Delta^{(\eta )-1}_{\mu\nu} (q)\ =\ q^2\, \Big( -g_{\mu\nu}
\, +\, \frac{q_\mu q_\nu}{q^2}\, -\, \alpha 
\frac{\eta_\mu \eta_\nu}{\eta^2}\, \Big)\, ,
\end{equation}
with $\alpha = (q\cdot\eta)^2/(q^2\eta^2)$ and $\xi =1$. This leads to the 
propagator
\begin{equation}
\label{etaDel}
\Delta^{(\eta )}_{\mu\nu} (q)\ =\ \frac{1}{q^2}\, \Big[ -\, g_{\mu\nu}\, +\, 
\frac{q_\mu \eta_\nu + q_\nu \eta_\mu}{q\cdot \eta}\, 
-\, \beta\,  \frac{q_\mu q_\nu}{(q\cdot\eta)^2}\, \Big]\,  ,
\end{equation}
where $\beta = (1+1/\alpha )\eta^2$. Within the framework of the GPT
in these gauges, we have to decompose the tri-gauge vertex 
$\Gamma_{\mu\nu\lambda}(q,p,k)$ as
\begin{equation}
\label{etaGPT}
\Gamma_{\mu\nu\lambda}(q,p,k)\ =\ \Gamma^{(\eta )}_{\mu\nu\lambda}(q,p,k)
\, +\, \Gamma^{P(\eta)}_{\mu\nu\lambda}(q,p,k)\, ,
\end{equation}
so that 
\begin{equation}
\label{etaWI}
\frac{1}{g}\, q^\mu\, \Gamma^{(\eta )}_{\mu\nu\lambda}(q,p,k)\ =\ 
\Delta^{(\eta )-1}_{\nu\lambda}(p)\, -\, 
                         \Delta^{(\eta )-1}_{\nu\lambda}(k)\, .
\end{equation}
A natural solution to Eqs.\ (\ref{etaGPT}) and (\ref{etaWI})
may be given by
\begin{eqnarray}
\label{eta3g}
\Gamma_{\mu\nu\lambda}^{(\eta )}(q,p,k) & =& g\Big[\,
(p-k)_\mu g_{\nu\lambda}\, +\, (k-q)_\nu g_{\mu\lambda}\, +\,
(q-p)_\lambda g_{\mu\nu}\nonumber\\
&&-\, \frac{\eta_\mu\eta_\nu\eta_\lambda}{(\eta^2)^2}\,
[(k-p)\cdot\eta]\, \Big]\, ,\\
\label{eta3gP}
\Gamma_{\mu\nu\lambda}^{P(\eta )}(q,p,k) &=& g\, 
\frac{\eta_\mu\eta_\nu\eta_\lambda}{(\eta^2)^2}\,
[(k-p)\cdot\eta]\, .
\end{eqnarray}
Employing the identity
\begin{equation}
\label{etaProp}
\eta^\mu\, \Delta^{(\eta )}_{\mu\nu} (q)\ =\ 
-\, \frac{(\eta^2)^2\, q_\nu}{(q\cdot\eta)^3}\, ,
\end{equation}
we can readily see that $\Gamma_{\mu\nu\lambda}^{P(\eta )}(q,p,k)$ contains
{\em generalized} pinching momenta, {\em viz.}
\begin{equation}
\label{etaGamP}
\frac{1}{g}\, \Gamma_{\mu\nu\lambda}^{P(\eta )}(q,p,k)\, 
\Delta^{(\eta )\nu}_{\rho} (p)\, \Delta^{(\eta )\lambda}_{\sigma} (k)\ =\
\eta_\mu k_\sigma p_\rho\, \frac{(\eta^2)^2 [(k-p)\cdot\eta ]}{
(p\cdot\eta )^3 (k\cdot\eta )^3}\, .
\end{equation}

In these non-covariant gauges, it can be shown that ghosts decouple
from $S$-matrix elements completely in the DR \cite{JCT1}. To give
an example, we consider the ghost contribution to the gluon 
self-energy. The interaction Lagrangian containing the ghosts, $c^a_g$, may
be derived from
\begin{equation}
\label{etaghost}
{\cal L}_{ghost}\ =\ c^{a\dagger }_g\, \frac{\delta G^a [A]}{\delta\theta^b}
\, c^b_g\, ,
\end{equation} 
by calculating the response of the gauge-fixing condition $G^a[A]$ in Eq.\ 
(\ref{etaGF}) under an infinitesimal gauge transformation of the field
$A^a_\mu$, {\em i.e.},
\begin{equation}
A^a_\mu \  \to\ A^a_\mu\, -\, \frac{1}{g}\, \partial_\mu\theta^a\,
+\, f^{abc}\theta^b A^c_\mu\, .
\end{equation}
In this way, we find the ghost propagator, 
\begin{equation}
\label{Dab}
D^{ab} (q)\ =\ \frac{\delta^{ab}\, \eta^2}{(q\cdot\eta )^2}
\end{equation}
and the gluon-ghost-ghost coupling, $A^a_\mu(q)-c^b_g(p)-c^c_g(k)$,
\begin{equation}
\label{etaS}
S^{abc}_\mu(p,k)\ =\ -\, g\, f^{abc} \frac{(k\cdot\eta )}{\eta^2}\, \eta_\mu
\, ,
\end{equation}
with $q+p+k=0$. With the aid of the Feynman rules in Eqs.\ 
(\ref{Dab}) and (\ref{etaS}), it is straightforward to show the vanishing
of the ghost loop in the gluon self-energy. Indeed,
one has
\begin{eqnarray}
\label{etaPighost}
\Pi^{(ghost)}_{\mu\nu} (q) &\propto& \eta_\mu\eta_\nu\, \int
d^nk\, \frac{1}{(k\cdot \eta )^2}\, [(k+q)\cdot\eta   ]\, 
\frac{1}{[(k+q) \cdot \eta ]^2}\, (k\cdot \eta )\nonumber\\
&=& \eta_\mu\eta_\nu\, \int d^nk\, \frac{1}{(k\cdot\eta  )
[(k+q) \cdot \eta ]}\nonumber\\
&=& \frac{\eta_\mu\eta_\nu}{q\cdot \eta }\, \int d^nk\, \Big[\,
\frac{1}{k\cdot \eta }\, -\, \frac{1}{(k+q) \cdot \eta  }\, \Big]\, .
\end{eqnarray}
The last integral in Eq.\ (\ref{etaPighost}) vanishes through a shift of the
loop-momentum variable to the origin. By analogy, one can show that ghosts do
not contribute to three-gluon, four-gluon, {\em etc.}, vertices at one loop. 

Taking the afore-mentioned decoupling property of the ghosts into account,
we find that the effective GPT self-energy in non-covariant gauges is given by
\begin{equation}
\label{etaPTQCD}
\widehat{\Pi}^{(\eta )}_{\mu\nu} (q)\ =\ 
       \frac{c_A}{4}\, \int\, \frac{d^nk}{i(2\pi )^n}\,
\Big[\, \Delta^{(\eta )}(p) \Delta^{(\eta )}(k)\, 
\Gamma^{(\eta )}_\mu (q,p,k)\Gamma_\nu^{(\eta )}(q,p,k)\,
+\, T^{(tad.)}_{\mu\nu}\, \Big]\, ,
\end{equation}
where $T^{(tad.)}_{\mu\nu}$ is a tadpole contribution having the
form
\begin{equation}
\label{eta-tad}
\int d^nk\, T^{(tad.)}_{\mu\nu}\ \propto \   \eta_\mu\eta_\nu\, 
\int d^nk\, \frac{[(k-p)\cdot\eta]^2}{(k\cdot\eta )^2 (p\cdot\eta )^2}\ =\ 
\eta_\mu\eta_\nu\, \int d^nk\, \Big[\, \frac{1}{(k\cdot\eta )^2}\, +\, 
\frac{1}{(p\cdot\eta )^2}\, \Big]\, ,
\end{equation}
which vanishes in DR.
In addition, we obtain the same WIs with those of the GPT in the covariant
gauges and the PT, {\em i.e.},
\begin{eqnarray}
\label{etaWid}
q^\mu\, \widehat{\Gamma}^{(\eta )}_\mu (q,p_1,p_2) & = & 
g \Big[\, \widehat{\Sigma}^{(\eta )}(\not\! p_1)\,
                 -\, \widehat{\Sigma}^{(\eta )}(\not\! p_2 )\, \Big]\, ,\\
\label{etatr}
q^\mu\, \widehat{\Pi}^{(\eta )}_{\mu\nu} (q) &=& 0\, ,               
\end{eqnarray}
where $\widehat{\Sigma}^{(\eta )}(\not\! p)$ is the GPT quark self-energy in
the corresponding gauge. This self-energy is easily determined by taking the
gluon propagator in the loop in the non-covariant form of Eq.\ (\ref{etaDel}).
It should be stressed that the transversality identity in Eq.\ (\ref{etatr})
does not reassure that $\widehat{\Pi}^{(\eta )}_{\mu\nu} (q)$ is only 
proportional to $t_{\mu\nu} (q)$ in general. In non-covariant gauges, there
can exist another Lorentz structure \cite{JCT1} having the transversality 
property of Eq.\ (\ref{etatr}), which is given by 
\begin{equation}
\label{Nmunu}
n_{\mu\nu}(q,\eta )\ =\ [(q\cdot\eta )q_\mu\, -\, q^2\eta_\mu]\, 
[(q\cdot\eta )q_\nu\, -\, q^2\eta_\nu ]\, .
\end{equation}
As a result, the GPT self-energy can generally be expressed as follows:
\begin{equation}
\widehat{\Pi}^{(\eta )}_{\mu\nu} (q)\ =\ t_{\mu\nu} (q)\, 
\widehat{\Pi}^{(\eta )}_T (q)\, +\, 
n_{\mu\nu}(q,\eta )\, \widehat{\Pi}^{(\eta )}_N (q)\, .
\end{equation}
Beyond one loop, these gauges may pose some computational difficulties
\cite{JCT2}, since the higher-order correlation functions will contain
unphysical poles of the kind $1/(k\cdot\eta )$. Nevertheless, at one
loop, Mandelstam--Leibbrandt prescriptions for regularization of these
poles can lead to meaningful results \cite{GL&SM}.

Another and, perhaps, more familiar form of the gauge-fixing condition in 
non-covariant gauges is \cite{Axial,JCT1}
\begin{equation}
\label{eta1GF}
G^a[A]\ =\ \frac{1}{(\eta^2 )^{1/2}}\,  \eta^\mu A^a_\mu \, .
\end{equation}
The inverse gluon propagator in this gauge for any value of $\xi$ is given by
\begin{equation}
\label{eta1inv}
\Delta^{A-1}_{\mu\nu}(q)\ =\ q^2 t_{\mu\nu}(q)\, -\, 
                                    \frac{\eta_\mu \eta_\nu}{\xi\, \eta^2}\, ,
\end{equation}
which yields 
\begin{equation}
\label{eta1Del}
\Delta^{A}_{\mu\nu}(q)\ =\ \Big[\, -\, g_{\mu\nu}\, +\, 
\frac{q_\mu\eta_\nu + q_\nu \eta_\mu}{q\cdot\eta}\, -\,
\frac{\eta^2 q_\mu q_\nu}{(q\cdot\eta )^2}\, -\, \xi\, \frac{\eta^2 q^2 q_\mu 
q_\nu}{(q\cdot\eta )^2}\, \Big]\, \frac{1}{q^2}\ .
\end{equation}
For $\xi\not = 0$, the propagator in this class of gauges shows a bad
high-energy unitarity behaviour coming from the $\xi$-dependent term in Eq.\
(\ref{eta1Del}), which will affect multiplicative renormalization. The
characteristic feature of these gauges is the {\em complete} absence of a
pinching tri-gauge term, $\Gamma^{P,A}_{\mu\nu\lambda}(q,p,k)$, within the
framework of the GPT. It is easy to see that 
\begin{equation}
\label{eta1WI}
\frac{1}{g}\, q^\mu\, \Gamma_{\mu\nu\lambda} (q,p,k)\ =\ 
\Delta^{A-1}_{\nu\lambda}(p)\, -\, \Delta^{A-1}_{\nu\lambda}(k)\, .
\end{equation}
Taking Eq.\ (\ref{3gWI}) into account, we observe the presence of an extra
freedom in the RHS of Eq.\ (\ref{eta1WI}). To be specific, one can always
add to the inverse propagator $U^{-1}_{\mu\nu}(q)$ in the unitary gauge a
symmetric $q^2$-independent tensor, such as $\eta_\mu\eta_\nu/(\xi\eta^2)$
in Eq.\ (\ref{eta1inv}), without violating the WI of Eq.\ (\ref{eta1WI}).
Even though the whole algorithm of the GPT may be trivial in this case, the
absence of generalized pinching parts in the tri-gluon coupling can account
for the fact that the WIs in Eqs.\ (\ref{etaWid}) and (\ref{etatr}) are
automatically satisfied in this gauge. The latter will be valid for any
one-loop multi-point correlation function in QCD \cite{JMC,JMC_JP}.

Our considerations can equally well carry over to general Coulomb gauges,
which arise from the gauge-fixing condition,
\begin{equation}
\label{etaCoul}
G^a [A]\ =\ \Big( -g^{\mu\nu}\, +\, \frac{\eta^\mu\eta^\nu}{\eta^2} \Big)\, 
\partial_\mu A^a_\nu\, ,
\end{equation}
with $\eta^\mu\ =\ (1,0,0,0)$. Again, one can evaluate effective GPT
two-point, three-point, {\em etc.}, correlation functions, using the
extended PT algorithm outlined above.

In Section 2.1, it has become apparent that it exists a connection between the
GPT and the BFM in covariant gauges. One may therefore attempt to investigate
if there is a similar analogue for the GPT in non-covariant gauges. This will
be our main concern in the next section. 

\setcounter{equation}{0}
\section{Background field method in general gauges}
 
First, we shall briefly review the main features of the BFM in pure 
Yang-Mills theories, such as quark-less QCD with $N$ colours. For
more details, the reader is referred to \cite{Abbott}. Then, we shall
consider the BFM in general non-covariant gauges and compare our results
for the two-point Green's functions with those obtained by the GPT in the
respective gauge.

The BFM relies on the linear expansion of the gauge field, $A^a_\mu$,
about the background field, $\hat{A}^a_\mu$, which amounts to replacing
\begin{equation}
\label{BFMdec}
A^a_\mu\ \to \ \hat{A}^a_\mu\, +\, A^a_\mu\, ,
\end{equation}
in the classical Yang-Mills Lagrangian
\begin{equation}
\label{BFMYM}
{\cal L}_{YM}[\hat{A} + A ]\ =\ -\, \frac{1}{4}\, 
    F^a_{\mu\nu}[\hat{A} + A] \, F^{a,\mu\nu}[\hat{A} + A] \, ,
\end{equation}
where the field strength tensor, $F^a_{\mu\nu}[Q]$, of a Yang-Mills
field, $Q^a_\mu$, is defined as usual by
\begin{equation}
\label{Fmunu}
F^a_{\mu\nu}[Q]\ =\ \partial_\mu Q^a_\nu\, -\,
\partial_\nu Q^a_\mu\, +\, gf^{abc}Q^b_\mu Q^c_\nu\, .
\end{equation}
Adopting 't-Hooft's formulation in Ref.\ \cite{BFM}, it is not necessary to
assign a source term to $\hat{A}$, as only the field component $A$ gets
quantized. In fact, the quantum field $A$ is the integration variable
in the generating functional
\begin{equation}
\label{BFMZ}
Z [J,\hat{A} ]\ =\ \int [dA]\, 
      \mbox{det}\Big[ \frac{\delta G^a}{\delta\theta^b}\Big]\,
      \exp \Big[i\int d^4x \Big({\cal L}_{YM}-\frac{1}{2\xi_Q}(G^a)^2
      + J^a_\mu A^{a,\mu} \Big)\Big]\, .
\end{equation}      
In Eq.\ (\ref{BFMZ}), $G^a$ is the gauge-fixing condition and 
$\delta G^a/\delta\theta^b$ is its derivative under the infinitesimal
gauge transformation of the quantum field $A^a_\mu$,
\begin{equation}
\label{A-gtrafo}
\delta A^a_\mu\ =\ -\, \frac{1}{g}\, \partial_\mu\theta^a\, +\,
f^{abc}\theta^b (\hat{A}^c_\mu + A^c_\mu )\, .
\end{equation}

One of the main advantages of the BFM is that one can maintain gauge
invariance in $Z[J,\hat{A}]$ with respect to the background field
$\hat{A}^a_\mu$. In covariant gauges, one usually chooses the background
field gauge-fixing condition 
\begin{equation}
\label{BFMxiQ}
G^a [\hat{A},A]\ =\ \partial_\mu A^{a,\mu}\, +\, 
gf^{abc}\hat{A}^b_\mu A^{c,\mu}\ .
\end{equation}
With the gauge-fixing condition (\ref{BFMxiQ}), one can show that
$Z[J,\hat{A}]$ is invariant under the infinitesimal transformations
\cite{Abbott}: 
\begin{eqnarray}
\label{BFMtrafo1}
\delta\hat{A}^a_\mu & =& -\, \frac{1}{g}\, \partial_\mu \hat{\theta}^a\, +\, 
f^{abc}\hat{\theta}^b\hat{A}^c_\mu \ ,\\
\label{BFMtrafo2}
\delta J^a_\mu &=& -\, f^{abc}\hat{\theta}^b J^c_\mu\ .
\end{eqnarray}
In Eqs.\ (\ref{BFMtrafo1}) and (\ref{BFMtrafo2}), we have denoted the
infinitesimal parameter of the gauge transformation of $\hat{A}^a_\mu$
by $\hat{\theta}^a$. The parameter $\hat{\theta}^a$ should be regarded
independent of that appearing in the gauge transformation of the quantum
field $A^a_\mu$ in Eq.\ (\ref{A-gtrafo}). In order to prove that
\begin{equation}
\label{BFMinv}
Z [J,\hat{A}]\ =\ Z [J+\delta J, \hat{A}+ \delta\hat{A}]\, ,
\end{equation}
it is very helpful to make the following orthogonal transformation of the
integration variable $A^a_\mu$ in the vector space spanned by the generators
of the gauge group SU$(N)$:
\begin{equation}
\label{AtoA'}
A^a_\mu \ \to \ A'^a_\mu\ =\ A^a_\mu \, -\, f^{abc}\hat{\theta}^b A^c_\mu\ =\
O^{ab}(\hat{\theta})A^b_\mu\, .
\end{equation}
$O^{ab}(\hat{\theta}) =\delta^{ab} - f^{axb}\hat{\theta}^x$ is an
orthogonal matrix representing a rotation by an infinitesimal amount
$\hat{\theta}$. Thus, the change of variables given in Eq.\ (\ref{AtoA'})
leaves the integration measure invariant. On the other hand, the gauge-fixing
term in Eq.\ (\ref{BFMxiQ}) transforms as 
\begin{equation}
\label{BFMGFtr}
G^a [\hat{A}+\delta\hat{A},A ]\ =\ G^a [\hat{A},A']\, +\, f^{abc}
\hat{\theta}^bG^c [\hat{A},A' ]\, =\, O^{T,ac}(\hat{\theta})
G^c [\hat{A},A']\, .
\end{equation}
As a result, the background field gauge invariance of the term proportional to
$(G^a)^2$ in Eq.\ (\ref{BFMZ}) is evident, since this term is manifestly
invariant under orthogonal rotations, given in Eq.\ (\ref{BFMGFtr}). This
property of $G^a$ will turn out to be very crucial, while extending the
gauge-fixing condition to non-covariant background field gauges. Finally, it
is straightforward to calculate the derivative relation 
\begin{equation}
\label{BFMvar}
\frac{\delta}{\delta \theta^b}\, G^a [\hat{A}+\delta\hat{A},
A^\theta (\hat{A}+\delta\hat{A}) ]\ =\ O^{T,ac}(\hat{\theta})\, 
\frac{\delta}{\delta \tilde{\theta}^d}\, G^c [\hat{A},
A'^{\tilde{\theta}} (\hat{A}) ]\, O^{db}(\hat{\theta} )\, ,
\end{equation}
where $\tilde{\theta}^a= O^{ab}(\hat{\theta})\theta^b$. In Eq.\ (\ref{BFMvar}),
we have explicitly indicated the dependence of the variation of the quantum
field $A$ on the background field $\hat{A}$ and the infinitesimal parameter
$\theta$. It is now obvious that $\mbox{det} [\delta G^a/\delta\theta^b]$ is
invariant under transformations in Eqs.\ (\ref{BFMtrafo1}) and
(\ref{BFMtrafo2}), since $\mbox{det}[O^{ab}(\hat{\theta})] =1$. This completes
our proof of the equality in Eq.\ (\ref{BFMinv}). 

It is now useful to define the generating functional,
\begin{equation}
\label{BFMW}
W[J,\hat{A}]\ =\ -i\ln Z[J,\hat{A} ]\, ,
\end{equation}
whose derivatives give rise to connected Green's functions. Finally,
by performing a Legendre transformation, we define
\begin{equation}
\label{BFMGam}
\Gamma [\bar{A},\hat{A} ]\ =\ W[J,\hat{A}]\, -\, 
\int d^4x J^a_\mu\, \bar{A}^{a,\mu}\, ,
\end{equation}
with
\begin{equation}
\label{Abar}
\bar{A}^a_\mu\ =\ \frac{\delta W[J,\hat{A}]}{\delta J^{a,\mu}}\ .
\end{equation}
$\Gamma [\bar{A},\hat{A} ]$ is the effective action generating
one-particle irreducible (1PI) Green's functions. From Eqs.\ 
(\ref{BFMtrafo1})--(\ref{BFMinv}), we readily find
\begin{equation}
\label{GamInv}
\Gamma [\bar{A},\hat{A}]\ =\ \Gamma [\bar{A}',\hat{A}+\delta\hat{A}]\, ,
\end{equation}
with $\bar{A}'^a_\mu = O^{ab}(\hat{\theta })\, \bar{A}^b_\mu$. This implies 
that
\begin{equation}
\label{Invariance}
\Gamma [0,\hat{A}]\ =\ \Gamma[0,\hat{A}+\delta\hat{A} ]\, .
\end{equation}
Equation (\ref{Invariance}) shows the invariance of the effective action,
$\Gamma [0,\hat{A} ]$, under gauge transformations of the background field
$\hat{A}^a_\mu$. This may be considered to be the most central result of the
BFM. In the BFM, the background field, $\hat{A}^a_\mu$, occurs in external
lines, since $\langle \bar{A}^a_\mu \rangle =0$, whereas the quantum field,
$A^a_\mu$, appears in the loops only, as it is the dynamical variable which is
integrated over in $Z[J,\hat{A}]$. Further details about the relation between
$\Gamma [0,\hat{A} ]$ and the conventional quantum action $\Gamma [\bar{A}]$
may be found in \cite{Abbott}. 

Employing the Feynman rules derived from the BFM in covariant gauges
\cite{Abbott}, one can easily see that the gluon self-energy and the
one-loop $\hat{A}q\bar{q}$ vertex function are exactly equal with those
found with the GPT in $R_\xi$ gauges ({\em cf.}\ Eq.\ (\ref{xiQPTQCD})).
Another important consequence of background field gauge invariance is the
great simplification of the renormalization in Yang-Mills theories
\cite{Abbott}. If $Z_A$ and $Z_g$ are the $\hat{A}$-wave-function and
coupling renormalization constants, respectively, BFM imposes the equality
$Z_g=Z^{-1}_A$. The effect of this QED-like relation is that the running of
the effective coupling constant $g(\mu )$ is entirely determined by the
coefficient factor, $b_1$, which multiplies the ultra-violet (UV) divergent
part of $\widehat{\Pi}^{(\xi_Q )}_{\mu\nu}(q)$, as it happens in QED. At
one loop, the value of this coefficient factor is $b_1=11c_A/3$
\cite{Abbott}.

It is now worthwhile to investigate if the gauge-fixing condition given in Eq.\
(\ref{BFMxiQ}) is conceivably the only possible. To address this question, we
should first notice how the covariant derivative, defined in the adjoint
representation as 
\begin{equation}
\label{Dmu}
D^{ab}_\mu [\hat{A} ]\ =\ \delta^{ab}\partial_\mu\, +\, 
      gf^{axb}\hat{A}^x_\mu\ ,
\end{equation}
transforms under the background-field gauge transformation 
(\ref{BFMtrafo1}). Specifically, we have
\begin{equation}
\label{Dmu-trafo}
D^{ab}_\mu [\hat{A} +\delta\hat{A}]\, A^b_\lambda\ =\ 
D^{ab}_\mu [\hat{A} ]\, A'^b_\lambda\, +\, 
                     f^{abc} \hat{\theta}^b D^{cd}_\mu [\hat{A} ]\, 
                                                          A'^d_\lambda\, =
O^{T,ax}(\hat{\theta})D^{xz}_\mu [\hat{A} ]\, A'^z_\lambda\, .
\end{equation}
Evidently, the gauge-fixing condition
\begin{equation}
\label{BFMeta}
G^a [\hat{A}]\ =\ \frac{\eta^\mu\eta^\nu}{\eta^2}\,
D^{ab}_\mu [\hat{A} ]\, A'^b_\nu\, ,
\end{equation}
leaves $Z[J,\hat{A}]$ invariant under background field gauge transformations.
In fact, in order to get from the conventional approach in covariant and
non-covariant gauges to the corresponding BFM quantized action, it is
sufficient to make the replacement $\delta^{ab}\partial_\mu\to D^{ab}_\mu
[\hat{A} ]$, in the gauge-fixing term of the former.  Thus, the BFM analogue of
the general Coulomb gauge in Eq.\ (\ref{etaCoul}) will be given by
\begin{equation}
\label{BFMCoul}
G^a_{(Coul.)}[\hat{A}]\ 
         =\ \Big( -g^{\mu\nu}\, +\, \frac{\eta^\mu\eta^\nu}{\eta^2} \Big)\,
D^{ab}_\mu [\hat{A} ]\, A^b_\nu\, ,
\end{equation}
with the four-vector $\eta^\mu =(1,0,0,0)$. Another acceptable form
for a gauge-fixing condition is the one already encountered in 
Eq.\ (\ref{eta1GF}), {\em i.e.},
\begin{displaymath}
G^a[A]\ =\ \frac{1}{(\eta^2 )^{1/2}}\,  \eta^\mu A^a_\mu \, ,
\end{displaymath}
which trivially satisfies Eqs.\ (\ref{BFMGFtr}) and (\ref{BFMvar}).
Consequently, the effective action quantized via the gauge-fixing condition
of Eq.\ (\ref{eta1GF}) possesses a background-field-gauge invariance
inherently \cite{KM}. As has also been noticed in \cite{KM}, the $n$-point
BFM correlation functions (for $n\ge 2$) will be identical to the Green's
function's derived from the usual approach in the same gauge. This fact is
consistent with our earlier observation of the complete absence of {\em
generalized} pinching parts in the tri-gauge coupling
$\Gamma_{\mu\nu\lambda}(q,p,k)$ through the WI of Eq.\ (\ref{eta1WI}),
which renders GPT trivial in these gauges. Clearly, linear combinations
\cite{KM} of Eqs.\ (\ref{eta1GF}), (\ref{BFMxiQ}) and (\ref{BFMeta}) will
constitute acceptable forms of gauge-fixing conditions within the BFM
applied to a {\em renormalizable} Yang-Mills theory.

Instead of studying the most general gauge-fixing condition mentioned
above, we can show that the results obtained in the BFM in the
non-covariant gauge (\ref{BFMeta}) are equal to those found with the GPT in
the corresponding gauge ({\em cf.}\ Eq.\ (\ref{etaGF})). Assuming $\xi =1$
for simplicity, the quantum gluon propagator, $\delta^{ab}\Delta^{(\eta
  )}_{\mu\nu} (q)$, and the ghost propagator, $D^{ab} (q)$, do not differ
from those calculated by the conventional method in Eqs.\ (\ref{etaDel})
and (\ref{Dab}), respectively. The same holds true for the vertex
$\hat{A}^a_\mu (q)-A^b_\nu (p)-A^c_\lambda (k)$, which equals
$f^{abc}\Gamma^{(\eta )}_{\mu\nu\lambda } (q,p,k)$ in Eq.\ (\ref{eta3g}).
Therefore, it is obvious that the one-loop $\hat{A}q\bar{q}$-coupling will
be the same in both approaches. To calculate the gluon self-energy in the
BFM in non-covariant gauges, we need the coupling of $\hat{A}$ to ghosts,
{\em i.e.}, $\hat{A}^a_\mu (q)-c^b_g(p)-c^c_g (k)$.  This is given by
\begin{equation}
\label{etaAcc} S^{(\eta ) abc}_\mu (p,k)\ =\ -\, gf^{abc}\,
\frac{(p-k)\cdot\eta }{\eta^2}\, \eta_\mu\ . 
\end{equation} 
The gluon vacuum polarization of $\hat{A}$ is equal to $\widehat{\Pi}^{(\eta
)}_{\mu\nu} (q)$, up to an extra contribution coming from the ghost
interaction in Eq.\ (\ref{etaAcc}). However, closer inspection on the ghost
loop reveals that this is actually the tadpole term $T^{(tad.)}_{\mu\nu}$ in
Eq.\ (\ref{eta-tad}), which is irrelevant in DR. 

Furthermore, we must remark that the equivalence established between GPT and 
BFM in axial gauges will persist, even if the gluon propagators are taken
in some other gauge, {\em e.g.}, covariant $R_\xi$ gauge. For the one-loop
example of quark-quark scattering presented in Section 2, one then finds
that box-graphs will produce self-energy- and vertex-like GPT terms
induced from propagators differences of the form (\ref{xiQdiff}). These
GPT terms will cancel against corresponding terms present in the one-loop
coupling $A_\mu q\bar{q}$ and the vacuum polarization, leading to a unique
result. The connection between GPT and BFM will be elaborated further on
in Section 4.

Finally, it may be worth commenting on the fact \cite{JCT1} that the UV
divergent part of $\widehat{\Pi}^{(\eta )}_{\mu\nu} (q)$ evaluated in gauges
(\ref{eta1GF}) and (\ref{BFMeta}) is in general proportional to $b_1=11c_A/3$ 
only in the pure axial-gauge limit $\xi\to 0$, although the WIs in Eqs.\
(\ref{etaWid}) and (\ref{etatr}) are satisfied. The reason is that
$\widehat{\Pi}^{(\eta )}_{\mu\nu} (q)$ may contain UV divergences proportional
to $n_{\mu\nu}(q,\eta )$. These UV infinities will eventually induce 
non-covariant counter-terms \cite{JCT1} of the gauge-invariant form 
$\eta^\mu \eta^\nu F^a_{\mu\lambda} F^{a,\lambda}_\nu $.

\setcounter{equation}{0}
\section{Connection between GPT and BFM}

In this section, we shall demonstrate the relation between the GPT and the BFM
in the non-covariant gauge (\ref{etaGF}) for $\xi =1$, by analyzing a typical 
quark-quark scattering, {\em e.g.}, $q\bar{q}\to q'\bar{q}'$. Recently,
analogous considerations based on BRS identities have been applied to show
that basic field-theoretical requirements necessary for a resummation
formalism are satisfied by the PT \cite{JP&AP2}. In this context, we wish to
briefly address the issue of how unique is the tri-gauge decomposition in
Eqs.\ (\ref{xiQGPT}) and (\ref{etaGPT}) within the framework of the GPT in
covariant and non-covariant gauges. Furthermore, we will present a way to
deduce the effective tree-level four-gluon vertex present in a PT one-loop
amplitude, in which two of the gluons are in the loop while the other two are
external. The effective four-gluon coupling can be isolated from a proper 
one-loop four-point function by resorting to the intrinsic property
\cite{JMC_JP} of the PT. The analytic result so-derived is found to be the
same with the respective vertex, $\hat{A}\hat{A}AA$, in the BFM for $\xi_Q=1$.
In the conventional PT, the four-gauge coupling does not contain any pinching
momenta.  However, one could define a `pinching' term by decomposing the
tree-level four-gluon vertex into two parts, so that the `pinching' and
non-`pinching' part separately satisfy the same BRS identity. 

Let us consider the one-loop transition amplitude ${\cal T}^{(2)}= 
\langle q'\bar{q}'|T^{(2)}|q\bar{q}\rangle $, where the superscript on $T$ 
will denote the order of expansion in powers of $g^2$. Following Ref.\ 
\cite{JP&AP2}, we will calculate the absorptive part of  ${\cal T}^{(2)}$.
Then, one may rely on the analyticity property of the $S$ matrix to obtain
its dispersive or Hermitian part. After carrying out all possible unitarity
cuts, one can see that the absorptive amplitude depends on two tree-level 
matrix elements, where the first involves quarks in the intermediate state, 
{\em i.e.}, $\langle q''\bar{q}''|T^{(1)}|q\bar{q} \rangle $, and the second 
gluons, {\em e.g.}, $\langle gg|T^{(1)}|q\bar{q}\rangle $. More  explicitly, 
we have
\begin{eqnarray}
\label{unit}
\frac{1}{2i}\ \langle q'\bar{q}' |\big( T^{(2)} - T^{(2)\dagger} \big)
|q\bar{q} \rangle &=& \frac{1}{2} \int dX_{\scriptstyle LIPS}
\Big[\,  \sum_{q''} \langle q'\bar{q}'|T^{(1)}|q''\bar{q}''\rangle
\langle q\bar{q}|T^{(1)}|q''\bar{q}''\rangle^*\nonumber\\
&&+\, \frac{1}{2} \langle q'\bar{q}'|T^{(1)}| gg \rangle
\langle q\bar{q} |T^{(1)}| gg \rangle^*\, \Big]\, ,
\end{eqnarray}
where $dX_{\scriptstyle LIPS}$ indicates the two-body Lorentz-invariant
phase-space integration measure. The factor $1/2$ in front of the gluonic
contribution on the RHS of Eq.\ (\ref{unit}) is statistical.  It is now
obvious that any gauge dependence of the tree-level quark-dependent
amplitudes, {\em e.g.}, $\langle q''\bar{q}''|T^{(1)}|q\bar{q} \rangle$, is
trivial. Thus, one has the freedom to choose the virtual gluon propagator in
an arbitrary gauge, and hence the non-covariant one in Eq.\ (\ref{etaDel}).
This fact should be contrasted with our earlier observation of the vanishing
of the GPT terms in any gauge for the QED-like part of the vertex 
$\Gamma_{1\mu}(q,p_1,p_2)$ in Fig.\ 3. 

The situation is different for the case of the two gluons in the intermediate
state, $\langle q\bar{q} |T^{(1)}| g (l_1) g (l_2) \rangle$. If we denote this
matrix element by ${\cal T}_{\mu\nu}$, we then have the WI
\begin{equation}
\label{BRSgg}
l_1^\mu l_2^\nu {\cal T}_{\mu\nu} \ =\ 0\, .
\end{equation}
Furthermore, ${\cal T}_{\mu\nu}$ is the sum of two amplitudes: 
${{\cal T}_s}_{\mu\nu}$, characterized by the presence of the three-gluon
vertex in the $s$-channel, and ${{\cal T}_t}_{\mu\nu}$, the remainder. 
Within the GPT in the non-covariant gauge defined in Eq.\ (\ref{etaGF}), we
split ${{\cal T}_s}_{\mu\nu}$ according to Eq.\ (\ref{etaGPT}) as follows: 
\begin{equation}
\label{Tdec}
{{\cal T}_s}_{\mu\nu}\ =\ {{\cal T}_s}^{(\eta )}_{\mu\nu}\, +\, 
{{\cal T}_s}^{P(\eta )}_{\mu\nu}\, ,
\end{equation}
where ${{\cal T}_s}^{(\eta )}_{\mu\nu}$ contains the effective tree-level
tri-gauge coupling $\Gamma^{(\eta )}_{\mu\nu\lambda}(q,p,k)$ in Eq.\ 
(\ref{eta3g}) and ${{\cal T}_s}^{P(\eta )}_{\mu\nu}$ its pinching counterpart
$\Gamma^{P(\eta )}_{\mu\nu\lambda}(q,p,k)$ in Eq.\ (\ref{eta3gP}).
In the ghost-free gauges under consideration, the polarization 
tensor of gluons is given by
\begin{equation}
\label{Pmunu}
P_{\mu\nu}(q,\eta)\ =\ \sum\limits_{\lambda =1,2}\, 
\varepsilon^{(\lambda)}_\mu(q)\, \varepsilon^{(\lambda)}_\nu(q)\ =\
-\, g_{\mu\nu}\, +\, \frac{\eta_\mu q_\nu + \eta_\nu q_\mu}{q\cdot \eta}\,
-\, \eta^2\, \frac{q_\mu q_\nu}{(q\cdot \eta)^2}\ .
\end{equation}
This result would also have been obtained, if we had applied Cutkosky rules to
the gluon propagator in Eq.\ (\ref{etaDel}) and set $q^2=0$. Omitting the LIPS
integral of the intermediate gluons for brevity, one has for the bosonic
contribution in Eq.\ (\ref{unit}) that 
\begin{eqnarray}
\label{Tbos}
\frac{1}{2i}\ \langle q'\bar{q}' |\big( T^{(2)} - T^{(2)\dagger} \big)
|q\bar{q} \rangle_{bos.} &=& 
\frac{1}{4} {\cal T}_{\mu\nu} P^{\mu\rho} (l_1,\eta)P^{\nu\sigma} (l_2,\eta) 
{\cal T}^*_{\rho\sigma }\nonumber\\
&=& {\cal M}^{(\eta )}\, +\, \delta {\cal M}\, ,
\end{eqnarray}
where ${\cal M}^{(\eta )}$ is the absorptive amplitude one obtains within
the GPT or BFM in the ghost-free gauge mentioned above,
\begin{equation}
\label{Teta}
{\cal M}^{(\eta )}\, =\, \frac{1}{4} P^{\mu\rho} (l_1,\eta)P^{\nu\sigma} 
(l_2,\eta)\, \Big[\, {{\cal T}_s}^{(\eta )}_{\mu\nu}
{{\cal T}_s}^{(\eta )*}_{\rho\sigma}\, +\, {{\cal T}_s}^{(\eta )}_{\mu\nu}
{{\cal T}_t}^{(\eta )*}_{\rho\sigma}\, +\, 
{{\cal T}_t}^{(\eta )}_{\mu\nu}{{\cal T}_s}^{(\eta )*}_{\rho\sigma}\,
+\, {{\cal T}_t}^{(\eta )}_{\mu\nu}{{\cal T}_t}^{(\eta )*}_{\rho\sigma}\, \Big]
\end{equation}
and $\delta {\cal M}$ is the would-be deviation
\begin{equation}
\label{deltaM}
\delta {\cal M}\ =\ -\, \frac{1}{4}\, P^{\mu\rho} (l_1,\eta)P^{\nu\sigma}
(l_2,\eta)\, \Big[\, -\, {{\cal T}_s}^{P(\eta )}_{\mu\nu}
{{\cal T}_s}^{P(\eta )*}_{\rho\sigma}\, +\, {{\cal T}_s}^{P(\eta )}_{\mu\nu}
{\cal T}^*_{\rho\sigma}\, 
+\, {\cal T}_{\mu\nu}{{\cal T}_s}^{P(\eta )*}_{\rho\sigma}\, \Big]\, .
\end{equation}
On account of the WI in Eq.\ (\ref{BRSgg}) and the fact that $\eta^\mu
P_{\mu\nu} (l,\eta)=0$, $\delta {\cal M}$ vanishes identically. This example
explicitly demonstrates the connection between the GPT and the BFM in the
non-covariant gauge (\ref{etaGF}). Following a line of similar arguments, one
can reach the same conclusion for the GPT and the BFM in covariant or in more
general gauges discussed in Section 3. 

It is now interesting to analyze briefly to what extend the splitting of the
three-gluon vertex, $\Gamma_{\mu\nu\lambda} (q,p,k )$, given in Eq.\ 
(\ref{xiQGPT}), is uniquely determined, provided the non-pinching part of
$\Gamma_{\mu\nu\lambda} (q,p,k )$ satisfies the WI in Eq.\ (\ref{xiQWI}).
For example, another possible expansion of $\Gamma_{\mu\nu\lambda} (q,p,k )$
in covariant gauges would be 
\begin{eqnarray}
\label{3gxiQbar}
\bar{\Gamma}^{(\xi_Q )}_{\mu\nu\lambda} (q,p,k)& =& g \Big\{\, (p-k)_\mu
\Big[ g_{\nu\lambda}\, +\, \frac{p_\nu p_\lambda - k_\nu k_\lambda}{\xi_Q
(p^2 -k^2 )}\, \Big]\, +\, (k-q)_\nu g_{\mu\lambda}\nonumber\\
&&+\, (q-p)_\lambda g_{\mu\nu}\, \Big\}\, ,
\end{eqnarray}
which obeys the same WI of Eq.\ (\ref{xiQWI}), {\em i.e.},
\begin{equation}
\label{xiQbarWI}
\frac{1}{g}\, q^\mu \bar{\Gamma}^{(\xi_Q )}_{\mu\nu\lambda} (q,p,k)\ =\ 
\Delta^{(\xi_Q)-1}_{\nu\lambda} (p)\, -\, \Delta^{(\xi_Q)-1}_{\nu\lambda} 
(k)\ .
\end{equation}
The {\em generalized} pinching part of $\Gamma_{\mu\nu\lambda} (q,p,k )$
would then be defined as
\begin{equation}
\label{3gxiQbarP}
\bar{\Gamma}^{P(\xi_Q )}_{\mu\nu\lambda} (q,p,k)\ =\
g\, (p-k)_\mu\, \frac{p_\nu p_\lambda - k_\nu k_\lambda}{\xi_Q
(p^2 -k^2 )}\, .
\end{equation}
However, the form of $\bar{\Gamma}^{(\xi_Q )}_{\mu\nu\lambda} (q,p,k)$ as
well as that of $\bar{\Gamma}^{P(\xi_Q )}_{\mu\nu\lambda} (q,p,k)$ can only
arise from non-local interactions within a Lagrangian. Even if adopting
this realization of the GPT, one is still able to construct a three-point
Green's function $\widehat{\bar{\Gamma}}^{\displaystyle {}_{(\xi_Q )}}_\mu
(q,p_1,p_2)$, which satisfies the QED-like WI in Eq.\ (\ref{xiQWid}). An
example of the kind is the Vilkovisky-DeWitt effective action \cite{VDW},
which also predicts a non-local tri-gluon vertex.  One might therefore
expect that this action would correspond to the GPT in a particular gauge.
An extensive analysis of the latter lies beyond the scope of our present
discussion. In this context, we remark that, within the GPT in the
non-covariant gauge (\ref{etaGF}), the splitting of $\Gamma_{\mu\nu\lambda}
(q,p,k)$ in Eq.\ (\ref{etaGPT}) appears not to admit a non-local solution
beyond the local one found in Eqs.\ (\ref{eta3g}) and (\ref{eta3gP}).

So far, we have focused our attention on the tri-gauge coupling,
$\Gamma_{\mu\nu\lambda} (q,p,k )$. In the PT, the four-gluon coupling, 
$A^a_\mu A^b_\nu A^c_\nu A^d_\lambda $, given by
\begin{eqnarray}
\label{4g}
\Gamma^{abcd}_{\mu\nu\lambda\rho} &=&
-\, ig^2\, \Big[\, f^{abx} f^{xcd}\, ( g_{\mu\lambda} g_{\nu\rho}\, -
\, g_{\mu\rho} g_{\nu\lambda} )\, +\, f^{dax} f^{xbc}\, ( g_{\mu\lambda} 
g_{\nu\rho}\, - \, g_{\mu\nu} g_{\lambda\rho} )\nonumber\\
&&+\, f^{acx} f^{xbd}\, ( g_{\mu\nu} g_{\lambda\rho}\, -\, g_{\mu\rho} 
g_{\nu\lambda} )\, \Big]\, ,
\end{eqnarray}
does not contain any pinching part, as opposed to $\Gamma_{\mu\nu\lambda}
(q,p,k )$. In the BFM in covariant gauges, the coupling
$\hat{A}^a_\mu\hat{A}^b_\nu A^c_\lambda A^d_\rho$ is different from
$\Gamma^{abcd}_{\mu\nu\lambda\rho}$ appearing in the classical Lagrangian.
In a general $\xi_Q$ gauge in the BFM, the four-gauge coupling
$\hat{A}^a_\mu\hat{A}^b_\nu A^c_\lambda A^d_\rho$ may be written as 
\cite{Abbott}
\begin{eqnarray}
\label{BFM4g}
\Gamma^{(\xi_Q) abcd}_{\mu\nu\lambda\rho} &=&
-\, ig^2\, \Big[\, f^{acx} f^{xbd}\, ( g_{\mu\nu} g_{\lambda\rho}\, -
\, g_{\mu\rho} g_{\nu\lambda}\, +\, \frac{1}{\xi_Q}\, g_{\mu\lambda} 
                                                  g_{\nu\rho} )\, \nonumber\\
&&+\, f^{dax} f^{xbc}\, ( g_{\mu\lambda} g_{\nu\rho} -\, 
g_{\mu\nu} g_{\lambda\rho}\, -\, \frac{1}{\xi_Q}\, g_{\mu\rho}g_{\nu\lambda} )
                                                                   \nonumber\\
&&+\, f^{abx} f^{xcd}\, ( g_{\mu\nu} g_{\lambda\rho}\, -\, g_{\mu\rho} 
g_{\nu\lambda} )\, \Big]\, ,
\end{eqnarray}
This might make one think that the connection established for the tri-gauge
coupling may get spoiled for the four-gluon vertex, especially when one
compares the proper one-loop PT four-point function $q\bar{q} A^a_\mu
A^b_\nu$, shown in Fig.\ 6, with the corresponding one obtained by the BFM.
To show that this relation still exists, we intend to isolate the effective
four-gluon vertex, $\Gamma^{F,abcd}_{\mu\nu\lambda\rho }$, from the graph
in Fig.\ 6(d), after including the relevant pinching contributions
originating from the one-loop transition amplitude $q\bar{q} A^a_\mu
A^b_\nu$. Consequently, this analysis presented here will equally carry
over to the GPT in general gauges.

Let us consider the 1PI four-point Green's function,
$q (p_1) \bar{q} (p_2)\to A^a_\mu (k_1) A^b_\nu (k_2)$ shown in Fig.\ 6, 
where the external quarks and gluons are taken to be on-shell. According to
the diagrammatic approach of the intrinsic PT \cite{JMC_JP}, we will only keep
PT terms that are akin to the graph in Fig.\ 6(d). For simplicity, we will
work in the Feynman-'t-Hooft gauge. To be specific, the effective vertex,
$\Gamma^{F,abcd}_{\mu\nu\lambda\rho }$, may be deduced from Fig.\ 6(b), by
including the pinching parts resulting from the two graphs in Fig.\ 6(a). Up
to overall factors, external quark spinors and gluon polarizations, we have 
\begin{eqnarray}
\label{fig6c}
\mbox{Fig.\ 6(c)} &\sim & \int\frac{d^nk}{k^2p^2 l^2}\,
\Gamma^{c,\lambda}\, S(\not\! p_1 -\not\! k)\, \Gamma^{d,\rho}\, \Big[ \, 
\Gamma^{F,adx}_{\mu\rho\sigma}(k_1,k,l)
              \Gamma^{F,bcx}_{\nu\lambda\sigma }(k_2,p,-l)\, \nonumber\\
&&+\, 
\Gamma^{F,acx}_{\mu\lambda\sigma}(k_1,p,l)
                   \Gamma^{F,bdx}_{\nu\rho\sigma}(k_2,k,-l)\, \Big]\, ,\\
\label{fig6d}
\mbox{Fig.\ 6(d)} &\sim & \int\frac{d^nk}{k^2p^2}\, 
\Gamma^{c,\lambda}\, S(\not\! p_1 -\not\! k)\, \Gamma^{d,\rho}\, 
\Big( \Gamma^{abcd}_{\mu\nu\lambda\rho}\, -\, 
\Gamma^{P,abcd}_{\mu\nu\lambda\rho}\, \Big)\, ,
\end{eqnarray}
where $\Gamma^a_\mu =\lambda^a \gamma_\mu$ and 
$S(\not\! p)=1/(\not\! p - m)$
are the coupling of a gluon field, $A^a_\mu$, to quarks and the quark
propagator at the tree level, respectively. Furthermore, the intrinsic
pinch contributions originating from the two graphs in Fig.\ 6(a) have the 
form
\begin{eqnarray}
\label{4gPT}
 \Gamma^{P,abcd}_{\mu\nu\lambda\rho} &=& i \Delta^{F,\alpha\beta}(l)\, 
\Big[\Gamma^{\tilde{P},adx}_{\mu\rho\alpha}(l)
              \Gamma^{\tilde{P},bcx}_{\nu\lambda\beta}(-l)\, +\,
\Gamma^{\tilde{P},acx}_{\mu\lambda\alpha}(l)
              \Gamma^{\tilde{P},bdx}_{\nu\rho\beta}(-l)\, \Big]\nonumber\\
&=& ig^2\, \big(\, f^{acx}f^{xbd} g_{\mu\lambda} g_{\nu\rho}\, -\,
f^{dax}f^{xbc} g_{\mu\rho} g_{\lambda\nu}\, \big)\, ,
\end{eqnarray}
where $\Delta^{F}_{\alpha\beta}(q)=-g_{\alpha\beta}/q^2$,
$\Gamma^{\tilde{P},abx}_{\mu\nu\lambda}(k) =  g f^{abx} k_\lambda $ is that
term of $\Gamma^P_{\mu\nu\lambda} (q,p,k)$ in Eq.\ (\ref{3gP}), whose momentum
gets contracted, as can be seen from Eq.\ (\ref{4gPT}). It is now easy to
verify that $\Gamma^{F,abcd}_{\mu\nu\lambda\rho} = 
\Gamma^{abcd}_{\mu\nu\lambda\rho} - \Gamma^{P,abcd}_{\mu\nu\lambda\rho}$ is
exactly the expression obtained from Eq.\ (\ref{BFM4g}) for $\xi_Q=1$.  In the
GPT in covariant gauges, one can derive the effective coupling, 
$\Gamma^{(\xi_Q ),abcd}_{\mu\nu\lambda\rho}$ from Eq.\ (\ref{4gPT}), by making
the following replacements:
\begin{eqnarray}
\Delta^F_{\alpha\beta}(l) &\to & \Delta^{(\xi_Q)}_{\alpha\beta}(l)
\, ,\nonumber\\ 
\Gamma^{\tilde{P},abx}_{\mu\nu\lambda}(l) &\to & 
\Gamma^{\tilde{P}(\xi_Q ),abx}_{\mu\nu\lambda}(l)\, =\,  g\,
f^{abx}\, \frac{1}{\xi_Q}\, l_\lambda\, . \nonumber
\end{eqnarray}
In this way, we find that the analytic expression for the non-`pinching' part
of the four-gluon coupling is identical to that obtained by the BFM in Eq.\
(\ref{BFM4g}). 

\begin{center}
\begin{picture}(420,200)(0,0)
\SetWidth{0.8}

\ArrowLine(0,180)(40,180)\Text(0,190)[l]{\small $q(p_1)$}
\ArrowLine(40,140)(0,140)
\ArrowLine(40,180)(40,140)\Text(0,130)[l]{\small $\bar{q}(p_2)$}
\Gluon(40,180)(80,180){3}{4}\Text(60,190)[]{\small $d,\rho\, (k)$}
                                           \LongArrow(55,175)(65,175)
\Gluon(40,140)(80,140){-3}{4}\Text(60,130)[]{\small $c,\lambda\, (p)$}
                                           \LongArrow(55,145)(65,145)
\Gluon(80,180)(120,180){3}{4}\Text(87,160)[l]{\small $x,\sigma\, (l)$}
                                           \LongArrow(75,155)(75,165)
\Gluon(80,180)(80,140){3}{4}\Text(130,190)[r]{\small $a,\mu\, (k_1)$}
                                           \LongArrow(105,175)(95,175)
\Gluon(80,140)(120,140){-3}{4}\Text(130,130)[r]{\small $b,\nu\, (k_2)$}
                                           \LongArrow(105,145)(95,145)
\Text(140,110)[]{\bf (a)}

\Text(140,160)[]{$+$}

\Text(165,160)[l]{{\small (crossed graph)}}

\Text(280,160)[]{$+$}

\ArrowLine(300,180)(340,180)
\ArrowLine(340,140)(300,140)
\ArrowLine(340,180)(340,140)
\Gluon(340,180)(370,160){3}{3}\Text(345,190)[l]{\small $d,\rho\, (k)$}
\Gluon(340,140)(370,160){-3}{3}\Text(345,130)[l]{\small $c,\lambda\, (p)$}
\Gluon(370,160)(400,140){3}{3}\Text(410,130)[r]{\small $b,\nu $}
\Gluon(370,160)(400,180){-3}{3}\Text(410,190)[r]{\small $a,\mu $}

\Text(350,110)[]{\bf (b)}

\ArrowLine(0,80)(40,80)\Text(0,90)[l]{\small $q(p_1)$}
\ArrowLine(40,40)(0,40)
\ArrowLine(40,80)(40,40)\Text(0,30)[l]{\small $\bar{q}(p_2)$}
\Photon(40,80)(78.5,80){3}{4}\Text(55,90)[]{\small $d,\rho\, (k)$}
                                           \LongArrow(55,75)(65,75)
\Photon(40,40)(78.5,40){-3}{4}\Text(55,30)[]{\small $c,\lambda\, (p)$}
                                           \LongArrow(55,45)(65,45)
\Gluon(82.5,80)(120,80){3}{4}\Text(87,60)[l]{\small $x,\sigma\, (l)$}
                                           \LongArrow(75,55)(75,65)
\Photon(80,78.5)(80,42.5){3}{4}\Text(130,90)[r]{\small $a,\mu\, (k_1)$}
                                           \LongArrow(105,75)(95,75)
\Gluon(82.5,40)(120,40){-3}{4}\Text(130,30)[r]{\small $b,\nu\, (k_2)$}
                                           \LongArrow(105,45)(95,45)
\BBoxc(80,80)(5,5)\BBoxc(80,40)(5,5)

\Text(140,10)[]{\bf (c)}

\Text(140,60)[]{$+$}

\Text(165,60)[l]{{\small (crossed graph)}}

\Text(280,60)[]{$+$}

\ArrowLine(300,80)(340,80)
\ArrowLine(340,40)(300,40)
\ArrowLine(340,80)(340,40)
\Photon(340,80)(368,62){3}{3}\Text(345,90)[l]{\small $d,\rho\, (k)$}
\Photon(340,40)(368,58){-3}{3}\Text(345,30)[l]{\small $c,\lambda\, (p)$}
\Gluon(372,58)(400,40){3}{3}\Text(410,30)[r]{\small $b,\nu $}
\Gluon(372,62)(400,80){-3}{3}\Text(410,90)[r]{\small $a,\mu $}

\BBoxc(370,60)(5,5)

\Text(350,10)[]{\bf (d)}

\end{picture}\\
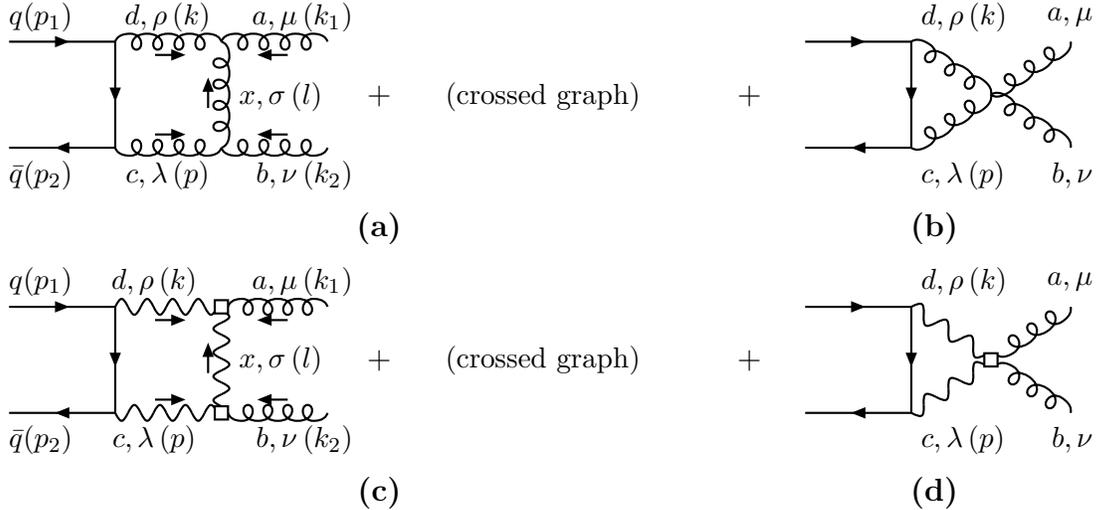
{\small {\bf Fig.\ 6:} Diagrammatic evaluation of the non-`pinching' part of
the four-gluon vertex.}
\end{center}

Analogous calculations may be carried out in the non-covariant gauge
(\ref{etaGF}). In this case, the gluon propagator is given in Eq.\
(\ref{etaDel}) for $\xi =1$. From Eq.\ (\ref{eta3gP}), the corresponding 
expression for $\Gamma^{\tilde{P}(\eta ),abx}_{\mu\nu\lambda}(l)$ is
identified to be
\begin{equation}
\label{etaPtild}
\Gamma^{\tilde{P}(\eta ),abx}_{\mu\nu\lambda}(l)\ =\ gf^{abx} 
\frac{\eta_\mu\eta_\nu\eta_\lambda}{(\eta^2 )^2}\, (l\cdot\eta )\, .
\end{equation}
After making the obvious substitutions of Eqs.\ (\ref{etaDel}) and
(\ref{etaPtild}) into Eq.\ (\ref{4gPT}), we obtain both for 
$\Gamma^{P(\eta ),abcd}_{\mu\nu\lambda\rho}$ and 
$\Gamma^{(\eta ),abcd}_{\mu\nu\lambda\rho}$ the following results:
\begin{eqnarray}
\label{eta4gP}
\Gamma^{P(\eta ),abcd}_{\mu\nu\lambda\rho} & = & 
ig^2\, \eta_\mu\eta_\nu\eta_\lambda\eta_\rho\, \frac{1}{(\eta^2)^2}\,
\big(\, f^{acx}f^{xbd} \, -\, f^{dax}f^{xbc}\, \big)\, ,\\
\label{eta4g}
\Gamma^{(\eta ),abcd}_{\mu\nu\lambda\rho} & = & 
\Gamma^{abcd}_{\mu\nu\lambda\rho}\, -\, 
\Gamma^{P(\eta ),abcd}_{\mu\nu\lambda\rho}\, ,
\end{eqnarray}
where $\xi =1$ and the analytic form of $\Gamma^{abcd}_{\mu\nu\lambda\rho}$ is 
given in Eq.\ (\ref{4g}). Equation (\ref{eta4g}) equals the expression one
would have obtained by calculating the Feynman rule of the coupling
$\hat{A}^a_\mu\hat{A}^b_\nu A^c_\lambda A^d_\rho$, directly from the BFM in the
gauge (\ref{BFMeta}). Once again, this nicely demonstrates the powerful
relation between the GPT and the BFM at the diagrammatic level.

Therefore, it makes sense to suggest that there is an effective pinching
part in the four-gluon vertex $A^a_\mu A^b_\nu A^c_\lambda A^d_\rho$, when
two of the gluons are in external lines, $A^a_\mu$ and $A^b_\nu$ say. 
For instance, in the GPT in covariant gauges, one may perform the 
decomposition (all momenta are incoming)
\begin{equation}
\label{4gxiQGPT}
\Gamma^{abcd}_{\mu\nu\lambda\rho}(q,p,k,r)\ =\ 
\Gamma^{(\xi_Q ),abcd}_{\mu\nu\lambda\rho}(q,p,k,r)\, +\,  
\Gamma^{P(\xi_Q ),abcd}_{\mu\nu\lambda\rho}(q,p,k,r)\, ,
\end{equation}
so that $\Gamma^{(\xi_Q ),abcd}_{\mu\nu\lambda\rho}$ satisfies the WIs
\begin{eqnarray}
\label{4gxiQWI1}
\frac{1}{ig}\, q^\mu \Gamma^{(\xi_Q),abcd}_{\mu\nu\lambda\rho}(q,p,k,r) &=&
f^{abx}\Gamma^{(\xi_Q),xcd}_{\nu\lambda\rho} (q+p,k,r)\, +\,
 f^{acx}\Gamma^{(\xi_Q),bdx}_{\nu\rho\lambda} (p,r,q+k)\nonumber\\ 
&&
+\,  f^{adx}\Gamma^{(\xi_Q),bcx}_{\nu\lambda\rho} (p,k,q+r)\, , \\
\label{4gxiQWI2}
\frac{1}{ig}\, p^\nu \Gamma^{(\xi_Q),abcd}_{\mu\nu\lambda\rho}(q,p,k,r) &=&
f^{bax}\Gamma^{(\xi_Q),xcd}_{\mu\lambda\rho} (p+q,k,r)\, +\,
 f^{bcx}\Gamma^{(\xi_Q),adx}_{\mu\rho\lambda} (q,r,p+k)\, \nonumber\\ 
&&
+\,  f^{bdx}\Gamma^{(\xi_Q),acx}_{\mu\lambda\rho} (q,k,p+r)\, ,
\end{eqnarray}
where the momenta $q$ and $p$ refer to the external gluons $A^a_\mu$
and $A^b_\nu$, respectively. Bear in mind that the first colour or Lorentz
index of $\Gamma^{(\xi_Q ),abc}_{\mu\nu\lambda}$ 
(or $\Gamma^{P(\xi_Q ),abc}_{\mu\nu\lambda}$) specifies which gluon is 
external, {\em e.g.}, $A^a_\mu$. Since $\Gamma^{abcd}_{\mu\nu\lambda\rho}
(q,p,k,r)$ obeys the same WIs in Eqs.\ (\ref{4gxiQWI1}) and (\ref{4gxiQWI2}),
involving the bare tri-gauge coupling $\Gamma^{abc}_{\mu\nu\lambda} (q,p,k)$,
the linear decomposition of the four-gluon vertex in Eq.\ (\ref{4gxiQGPT})
implies that the very same WIs should also be satisfied by their GPT
counterparts, {\em i.e.}, by $\Gamma^{P(\xi_Q ),abcd}_{\mu\nu\lambda\rho}
(q,p,k,r)$ and $\Gamma^{P(\xi_Q ),xcd}_{\nu\lambda\rho}(q+p,k,r)$, {\em etc}. 
This property may be considered as a consistency check of the GPT formalism. 

Some recent techniques devoted to the calculation of off-shell Green's
functions are based on the superstring formalism \cite{string}. At the
one-loop order, the superstring action naturally singles out the BFM. In other
approaches, the main concern has been to design new gauges that can simplify
diagrammatic analyses \cite{CSL}, in the same sense that Gervais-Neveu
\cite{GN} gauge reduces labor considerably in the calculation of multi-gluon
graphs in the tree approximation. Even though such methods may have some
relation with the GPT, our focal interest, however, is very different. Apart
from the link that we have established between the GPT and the BFM, our
primary aim is to formulate an algorithmic method, such as the GPT, which
gives rise to Green's functions in general gauges that satisfy Abelian-type
WIs, in line with resummation requirements \cite{JP&AP2}. The basic
diagrammatic rules governing the GPT in DR may be summarized as follows: 
\begin{itemize}
\item[1. ] Three-gluon vertices, in which one of the gluons is attached via an
external line, should be decomposed into two terms: the `pinching' and the
non-`pinching'. The non-`pinching' term is defined as that term that
satisfies the WI involving propagators in the given gauge, in which the GPT is
applied. For example, in covariant and non-covariant gauges, these WIs are
given in Eqs.\ (\ref{xiQWI}) and (\ref{etaWI}), respectively. The `pinching'
part gives rise to terms that should be added to lower $n$-point correlation
functions. As has been analyzed in this section, an analogous decomposition
may be considered for the four-gluon vertex as well, provided two of the
gluons are not in the loop. \item[2. ] Ghost couplings or ghost propagators of
a given gauge should not be modified in this algorithm. In particular, these
interactions are absent in ghost-free gauges. 
\item[3. ] Three- and four-gauge couplings in the loop, which do not have any
external gluons, are taken to be equal with $\Gamma^{abc}_{\mu\nu\lambda}
(q,p,k)$ in Eq.\ (\ref{3g}) and $\Gamma^{abcd}_{\mu\nu\lambda\rho}(q,p,k,r)$ in
Eq.\ (\ref{4g}), respectively.  The same holds true for the four-gauge
coupling, if three of the four gluons are in the loop.
\item[4. ] Within our algorithm, the difference of propagators can produce
residual pinch contributions. In fact, these residual pinch terms are
obtained, when one converts the gluon propagators in the loop, evaluated in an
arbitrary gauge, to the given gauge that the GPT is considered ({\em cf.}\
Eq.\ (\ref{xiQdiff})). 
\end{itemize}
These rules coincide with those of the usual PT at one loop in the
Feynman--'t-Hooft gauge \cite{JMC,JMC_JP}.  Nevertheless, beyond one loop,
these rules are inspired by the BFM. One may therefore expect that high-order
Green's functions will retain the desirable property of satisfying BFM WIs,
known from the PT at one loop.

\section{Conclusions}

It has been shown that Cornwall's PT \cite{JMC} can consistently be generalized
in such a way that the resulting Green's functions satisfy the very same WIs
known from the PT or the BFM. In Section 2, this extended version of the PT,
the GPT, is found to be very closely related to the BFM in covariant and
non-covariant gauges. However, contrary to the PT, the GPT Green's functions do
generally contain unphysical poles of the type, $1/(k\cdot\eta )^2$, in
non-covariant gauges. Moreover, the GPT and the BFM correlation functions in
the covariant gauges display $\xi_Q$-dependent thresholds in SSB theories
\cite{JP&AP2}. In this context, it should be stressed again that only the
conventional PT does not suffer from unphysical poles when unitarity cuts are
considered. In addition, conventional PT obeys a number of field-theoretical
requirements derived from resummation considerations \cite{JP&AP2}, thus
rendering it an appealing method for describing the dynamics of unstable 
particles at high-energy colliders \cite{JP&AP1}. 

The explicit connection between the GPT and the BFM is also demonstrated in a
typical quark-quark scattering in Section 4. The importance of our new
diagrammatic method formulated here, the GPT, may be seen from the fact that it
enables the design of new quantum actions at one loop and so provides a better
understanding of a possible link with the BFM. Conversely, the BFM in a
general gauge may be represented by the corresponding GPT at the $S$-matrix
level. Even though we have shown that such an one-to-one correspondence exists
for a wide class of covariant and non-covariant gauges, there may be some
cases, for which this relation could still fail and appropriate modifications
of the GPT algorithm and/or the BFM may be necessary. In the Vilkovisky-DeWitt
approach \cite{VDW}, for example, the gauge dependence of the usual effective
action is attributed to the choice of a coordinate system on an infinite
dimensional manifold of all field configurations. Thus, reparametrization
invariance of the field variables can be achieved by modifying the BFM
decomposition in Eq.\ (\ref{BFMdec}), supplemented by additional geometrical
restrictions. As a consequence, the effective Vilkovisky-DeWitt action
predicts non-local couplings coming from the affine connection. Nevertheless,
the GPT admits the treatment of such non-local interactions and can hence give
a unique chance to obtain a new insight between diagrammatic cancellations and
new gauges within the BFM. An analytic discussion of the latter may be given 
elsewhere. 

Beyond one loop, the PT and the GPT require an extensive study. Although the
algorithmic rules of the GPT given at the end of Section 4 may entail that the
above connection with the BFM is valid to all orders, it is the flexibility of
the former which allows us to believe that this solution may not be the only
possible one. Within the perturbation theory, the power of a diagrammatic
method, such as PT or GPT, lies in carrying out resummations of self-energies
inside the quantum loops in Yang-Mills theories. In fact, Cornwall's original
motivation has been based on that aspect, who has defined the PT by studying
specific three-loop graphs. This flexibility is not available in the BFM at the
moment, unless a variation of the loopwise expansion of the effective action
is to be invented. Even though we have not addressed the validity of the GPT in
SSB theories, the fact that the PT has successfully been applied to these
theories as well \cite{JPetal} makes one to conclude safely that the general
connection established in the present paper between the GPT and the BFM will
generally hold true. 

\vskip1.5cm
\noindent
{\bf Acknowledgments.} I wish to thank Joannis Papavassiliou for valuable
comments and for drawing my attention to Ref.\ \cite{BHJr}.

\end{document}